\documentclass[onecolumn,draftcls,12pt, twoside]{IEEEtran}
\pdfoutput=1
\usepackage[usenames, dvipsnames]{color}
\usepackage[cmex10]{amsmath} 
\usepackage{amssymb,amsthm}

\usepackage[nolist]{acronym}

\usepackage{enumerate}

\usepackage{subcaption}
\usepackage{calc}
\usepackage{cite}
\usepackage{graphicx}

\usepackage{epstopdf}

\usepackage{array,multirow}

\usepackage{cite}
\usepackage{tikz,pgfplots}
\usepackage{comment}

\usepackage{mathtools}

\usepackage[bookmarks=false]{hyperref}

\usepackage{amsmath} 

\newcommand{\mpr}{K}
\newcommand{\tf}{d}
\newcommand{\cri}{l}
\newcommand{\pgf}{Q}

\newcommand{\nuser}{n}

\newcommand{\ecri}{L}

\newcommand{\T}{\mathrm{T}}

\newcommand{\E}{\mathrm{E}}

\begin{acronym}
\acro{AP}{access point}
\acro{RA}{random access}
\acro{LT}{Luby Transform}
\acro{BP}{belief propagation}
\acro{i.i.d.}{independent and identically distributed}
\acro{IoT}{Internet of things}
\acro{MTC}{machine type communication}
\acro{PER}{packet error rate}
\acro{SIC}{successive interference cancellation}
\acro{URLLC}{ultra-reliable low latency communication}
\acro{PMF}{probability mass function}
\acro{CRI}{collision resolution interval}
\acro{PGF}{probability generating function}
\acro{CPGF}{conditional probability generating function}
\acro{EGF}{exponential generating function}
\acro{TGF}{transformed generating function}
\acro{rhs}{right-hand side}
\acro{lhs}{left-hand side}
\acro{CSA}{coded slotted ALOHA}
\acro{MPR}{multi-packet reception}
\acro{MTA}{modified binary-tree algorithm}
\acro{SICTA}{binary tree-algorithm with SIC}
\acro{BTA}{binary tree-algorithm}
\acro{MST}{maximum stable throughput}
\acro{MTA}{modified tree-algorithm}
\acro{CRP}{collision-resolution protocol}
\acro{CAP}{channel-access protocol}
\acro{IRSA}{Irregular Repetition Slotted ALOHA}
\acro{CSA}{Coded Slotted ALOHA}

\acro{r.v.}{random variable}
\end{acronym}

\title{
Tree-Algorithms with Multi-Packet Reception and Successive Interference Cancellation  
}

\author{\IEEEauthorblockN{\v Cedomir Stefanovi\' c\IEEEauthorrefmark{1}, Yash Deshpande\IEEEauthorrefmark{2}, H. Murat G\"ursu\IEEEauthorrefmark{3},  Wolfgang Kellerer\IEEEauthorrefmark{2} }\\
\IEEEauthorblockA{
\IEEEauthorrefmark{1}Department of Electronic Systems, Aalborg University, Denmark\\
\IEEEauthorrefmark{2}Chair of Communication Networks, Technical University of Munich, Germany \\
\IEEEauthorrefmark{3}Nokia Bell Labs, Munich, Germany\\
Email: cs@es.aau.dk, \{yash.deshpande,murat.guersu,wolfgang.kellerer\}@tum.de }}

\begin{document}

\maketitle


\thispagestyle{empty}
\pagestyle{empty}

\begin{abstract}
In this paper, we perform a thorough analysis of binary tree-algorithms with \ac{MPR} and \ac{SIC}, showing a number of novel results.
We first derive the basic performance parameters, which are the expected length of the collision resolution interval and the normalized throughput, conditioned on the number of contending users.
We then study their asymptotic behaviour, identifying an oscillatory component that amplifies with the increase in \ac{MPR}.
In the next step, we derive the throughput for the gated and windowed access assuming Poisson arrivals.
We show that for windowed access, the bound on maximum stable normalized throughput increases with the increase in \ac{MPR}.
This implies that investing in advanced physical capabilities, i.e., \ac{MPR} and \ac{SIC} pays off from the perspective of the medium access control algorithm.
\end{abstract}

\section{Introduction}
\label{sec:Intro}

In the last decade, there have been significant theoretical advances in the area of random-access protocols, instigated by the novel use-cases pertaining to the Internet of Things (IoT).
A typical IoT scenario involves a massive number of sporadically active users exchanging short messages.
The sporadic user activity mandates the use of random-access protocols, however, their use in massive IoT scenarios faces the challenge of an increased requirement for efficient performance.
Particularly, as the amount of exchanged data is low, the overhead of the random-access scheme should be minimal in order not to create a bottleneck in the overall communication setup.

A way to improve the performance of random-access protocols is to embrace the interference from the contending users.
Effectively, this is achieved by employing \ac{MPR}, enabled by the advanced capabilities of the physical layer (i.e., via the use of advanced signaling processing).
Tree-algorithms~\cite{capetanakis1979tree} and ALOHA~\cite{Abramson:ALOHA,R1975} are families of random-access protocols that by design suffer from collisions by contending users; as such, it is fitting to assume that protocols from these families will benefit from~\ac{MPR}.
Indeed, it was shown that \ac{MPR} improves the performance of slotted ALOHA, e.g.,~\cite{GVS1988,ZZ2012,GSP2015}.

Another well-explored line of research in the context of slotted ALOHA is the use of \ac{SIC} across slots.\footnote{Strictly speaking, \ac{SIC} is just another form of \ac{MPR}, and indeed many \ac{MPR} schemes rely on interference cancellation. In this paper, we use the term SIC to denote the application of interference cancellation across slots, while we assume that \ac{MPR} operates on individual slot basis.}
In this class of protocols~\cite{PSLP2014}, the users transmit multiple replicas of their packets on purpose. Decoding of a packet replica occurring in a singleton slot enables removal of all the other related replicas, potentially transforming some of the collision slots into singletons from which packets of other contending users can be decoded, and thus propelling new iterations of \ac{SIC}, etc.
The use of \ac{SIC} pushes the throughput performance significantly, asymptotically reaching the ultimate bound for the collision channel of 1~\rm{packet/slot}~\cite{L2011,Liva2015:CodedAloha}.
Finally, it was shown that a combination of \ac{MPR} and \ac{SIC} pushes the performance further than any of the two techniques separately~\cite{SPL2017}. 

A tree-algorithm based scheme exploiting \ac{SIC}, named SICTA, was proposed by Yu and Giannakis in~\cite{SICTA}.
It was shown that the \ac{MST} of SICTA reaches $\ln 2 \approx 0.693$~packet/slot, which is significantly better than the best performing variant of the algorithm without \ac{SIC}~\cite{V1986}.
Finally, the use of \ac{MPR} in tree-algorithms was analyzed in our recent work~\cite{SGDK2020}, where it was shown that~\ac{MPR} pushes the normalized\footnote{Normalized with respect to the assumed linear increase in physical resources to achieve \ac{MPR}.} \ac{MST} in the version of the scheme with the windowed access.

Motivated by the insights in \cite{SPL2017,SGDK2020}, in this paper we study the performance of tree-algorithms with $\mpr$-\ac{MPR} and \ac{SIC}; that is, we assume that the receiver is capable of successfully decoding any collision of up to and including $\mpr$ concurrent packet transmissions and can perform \ac{SIC} along the tree.
This is a simple, but widely used and insightful model that enables the assessment of the performance of multiple access algorithms.
We show a somewhat surprising fact that, for the gated access, the bound on the \ac{MST} (normalized with $\mpr$) decreases with $\mpr$, where this decrease is due to oscillations whose amplitude grows with $\mpr$.
On the other hand, in the case of the windowed access, the bound on normalized \ac{MST} increases with $\mpr$, surpassing the value of $\ln 2$. 
Specifically, the contributions of this paper are the following:
\begin{enumerate}
    \item We derive the expression for the expected length of a \ac{CRI} for \ac{BTA} with \ac{MPR} and \ac{SIC}, conditioned on the number of contending users $\nuser$, which is the prerequisite for the further analysis.
    \item We derive the asymptotic expressions for the expected \ac{CRI} length as $\nuser \rightarrow \infty$. 
    We also develop simple upper and lower bounds. 
    \item We investigate the bounds on the \ac{MST} for gated and windowed access method.
    \item We extend the derived results for \ac{BTA} with \ac{MPR} (without \ac{SIC}), preliminary investigated in \cite{SGDK2020}.
    \item Finally, we discuss the advantages and disadvantages of tree-algorithms with \ac{MPR} and \ac{SIC} in comparison with their slotted-ALOHA based analogues, e.g.,~\cite{L2011,SPL2017}.  
\end{enumerate}

The rest of the text is organized as follows.
In Section~\ref{sec:background} we state the background and briefly review the related work.
Section~\ref{sec:model} formulates the system model.
In Section~\ref{sec:analysis}, we derive the basic performance parameters, which are the expected \ac{CRI} and throughput, conditioned on the number of initially colliding users, followed by the derivation of their asymptotic values as well as upper and lower bounds.
Section~\ref{sec:arrivals} investigates the \ac{MST} performance of the scheme in the case of Poisson arrivals, both for the gated and the windowed access.
Section~\ref{sec:conclusion} concludes the paper with some results on tree algorithms with MPR (and no SIC) and a comparison with slotted-ALOHA based algorithms.


\section{Background and Related Work}
\label{sec:background}

\subsection{Tree Algorithms}
\label{sec:TA}

\begin{figure*}[t!]
  \centering
  \includegraphics[width=0.75\linewidth]{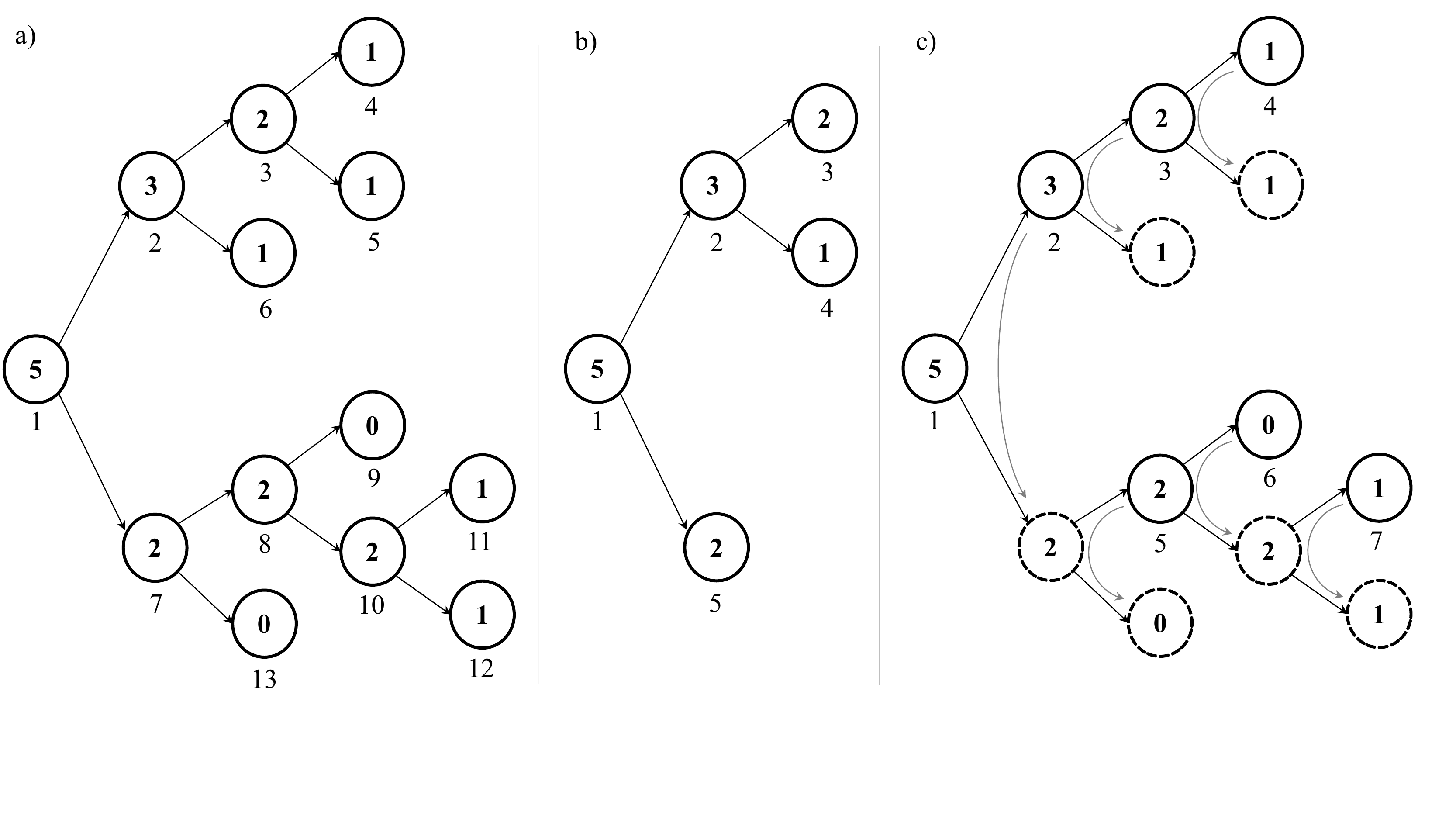}
  \caption{Illustrative examples of binary tree-algorithms. A node of the tree represents a slot, the number inside the node represents the number of users (i.e., packets) colliding in the slot, and the number beneath represents the sequence number of the slot. a) \ac{BTA} (the original version of the algorithm): It is assumed that initially 5 users collided, which progressively split in two groups until they become resolved (i.e., their packets decoded).
  b) BTA with 2 \ac{MPR}: The receiver is able to decode collisions of 2 or less packets, which reduces the number of slots required to resolve all users.
  c) Binary tree-algorithm with \ac{SIC}: The dashed nodes represent slots that are skipped, as the users belonging to the corresponding groups are resolved by cancelling the interference of the users resolved in the sibling node from the parent node, as indicated by the grey arrows.}
  \label{fig:tree-comparisons}
\end{figure*}

Tree-algorithms were introduced by Capetanakis~\cite{capetanakis1979tree}.
The key ingredient of tree-algorithms is the \ac{CRP} which is driven by the feedback sent by the receiver.
The basic variant of the \ac{CRP}, denoted as \ac{BTA}, operates as follows on a time-slotted multiple-access collision channel with feedback.
Assume that $\nuser$ users transmitted their packets in a slot:
\begin{itemize}
\item If $n=0$, the slot is idle, and the corresponding feedback is sent by the receiver.
\item If $n=1$, there is a single transmission in the slot (the slot is singleton), the packet is decoded (i.e., the user that transmitted the packet becomes resolved), which is acknowledged by the receiver.
\item If $\nuser > 1$, a collision occurs and the receiver sends the corresponding feedback, initiating the collision resolution.
The collided users split into two groups, e.g., group 0 and group 1; the decisions of which group to join are made uniformly at random and independently of any other user.
In the next slot, the users in group 0 transmit.
If the slot is idle (i.e., no user selected group 0) the users from group 1 transmit in the next slot.
If the slot is singleton, the packet in it is decoded and the users from group 1 transmit in the next slot.
Finally, if the slot is a collision slot, the users in group 0 split again in two groups and the procedure is recursively repeated.
In this case, the users in group 1 wait until all packets from users in group 0 become decoded (the information of which is obtained via monitoring the feedback).
\item The collision resolution ends when all $\nuser$ packets are decoded.
\end{itemize}
Fig.~\ref{fig:tree-comparisons}a) shows an example of the scheme.
In practice, the \ac{CRP} is combined with a \ac{CAP}, which specifies when the arriving (i.e., active) users can access the channel.
The basic variants of \ac{CAP} are the gated (also known as blocked), windowed and free access; the details about the former two are presented in Section~\ref{sec:arrivals}. 
It was shown that the \ac{MST} throughput of \ac{BTA} with the gated access is 0.346~packet/slot~\cite{capetanakis1979tree}.

This initial work inspired a number of research works on tree-algorithms.
Here we mention the modified tree-algorithm, which omits the slots that are certain to repeat the immediate previous collision (e.g., slot 10 in Fig.~\ref{fig:tree-comparisons}a) would be skipped), boosting the \ac{MST} with the gated access to 0.375~packet/slot~\cite{massey1981collision}.
Another significant improvement can be obtained by using the \ac{BTA} in the windowed access framework, which boosts the \ac{MST} to 0.429 packet/slot.

A further modification to the framework is made by considering $d$-ary splitting, a generalization in which the colliding users split into $d \geq 2$ groups.
It was shown in~\cite{MF1985} that the ternary tree-algorithms with biased splitting (biased meaning that the probabilities of choosing a group are not uniform over the groups) are the optimal choice. 
In this respect, a variant of the MTA scheme with clipped access (a modification of the windowed access) introduced in~\cite{V1986} is the best performing scheme in conventional setups (i.e., without \ac{MPR} and \ac{SIC}), achieving the \ac{MST} of 0.4878 packet/slot.

Supplementing tree-algorithms with $\mpr$-\ac{MPR} capability has the potential to improve their performance, as demonstrated in Fig.~\ref{fig:tree-comparisons}b).
However, to the best of our knowledge, works studying the impact of $K$-\ac{MPR} capability on tree-algorithms are scarce.
We mention the work deriving an upper and lower bound on the \ac{MST} (therein referred to as the capacity) for $K$-\ac{MPR} tree algorithm~\cite{TsyMikLik83}.
The work in~\cite{LikPloSha93} proposes a $K$-\ac{MPR} tree algorithms with an adaptive form of windowed access, where a part of the subsequent arrival window is added to the one being currently resolved, depending on the outcomes of the collision resolution.
The work in \cite{gau2011tree} analyzes \ac{MPR} in a tree-algorithm with continuous arrivals with a small number of users in the system (of the order of 10), proposing a transmission strategy that guarantees stability.
Finally, the paper~\cite{SGDK2020} performs the analysis of \ac{BTA} with $K$-\ac{MPR} in the standard windowed access setup; it is interesting to note that, as $\mpr$ grows, the \ac{MST} of the scheme becomes increasingly close to the lower bound on capacity derived in~\cite{TsyMikLik83}. 

A modification of the original scheme that employs \ac{SIC}, denoted as SICTA, was introduced in~\cite{SICTA}.
In SICTA, the receiver stores collision slots; once a packet becomes decoded in a slot occurring after a split has been performed, the receiver removes its replica from the previous collision slot(s) using \ac{SIC}, potentially instigating decoding of new packets and replica removal along the tree.
Fig.~\ref{fig:tree-comparisons}c) shows an example of SICTA; obviously, the use of \ac{SIC} enables skipping of the slots laying on the lower branches of the tree. 
The \ac{MST} of binary SICTA is 0.693, which is a huge improvement over \ac{MTA}.
However, we note that the results presented in~\cite{SICTA} for $\tf > 2$ do not hold - related insights on the performance analysis of $\tf$-ary SICTA, where $\tf > 2$, are presented in a separate work of ours~\cite{d-ary}.

We also mention the work presented in~\cite{GSP2018}, proposing a hybrid multiple-access scheme in which the user signatures are resolved via a $\mpr$-\ac{MPR} tree-algorithm (both with and without \ac{SIC}) and the user data via a polling mechanism.
The analysis of the tree-algorithm-based part of the scheme is basic, only providing bounds on the expected length of \ac{CRI} given the number of colliding users, and the main performance parameter is the net-rate taking into account the user data, which represents a dominant part of users' transmissions.


Finally, for the sake of completeness, we mention the variant of tree-algorithms with the free access~\cite{MF1985}, in which the users are free to access the channel as soon as they experience a packet arrival.
The \ac{MST} performance of the ternary \ac{MTA} with the free access falls between the one of the gated and one of the windowed access~\cite{MF1985}.
Further, the performance of tree-algorithms with the free access and with SIC or with MPR was investigated in~\cite{PB2009}.
However, we note that the approach to their analysis differs from the one presented in the paper, and this class of tree-algorithms is out of the paper scope.


\subsection{Multi-Packet Reception}

Research and design of multiple-access schemes that enable multi-packet reception has a long history; the canonical examples being CDMA, or Zadoff-Chu-preamble-based random-access used in 3GPP standards from LTE onwards~\cite{3GPPTS36.321}. 
There are also coding techniques specifically designed for this purpose -- we mention the $\mpr$-out-of-$\nuser$ coding for multiple-access channels, see~\cite[Chapters 2 and 3]{macbook-original},\cite{OP2017}. 

Some general models of \ac{MPR} capability from the perspective of random-access protocol can be found in, e.g., \cite{Ghez1988, TZM2001}.
In this paper, we adopt the following model:
\begin{enumerate}[(i)]
\item if there are up to and including $\mpr$ packets colliding in a slot, all packets are successfully decoded, and
\item if the number of colliding packets in a slot is greater than $\mpr$, no packet can be successfully decoded.
\end{enumerate}
This model can be understood as an extension of the collision channel model (the default channel model for the assessment of random-access algorithms).
Specifically, it can be referred to as the \emph{$\mpr$-collision channel}.

The works studying random-access protocols with \ac{MPR} typically abstain from modelling the investments required at the physical layer in order to enable the \ac{MPR}.  
In this paper, we assume that the $\mpr$-\ac{MPR} capability requires $\mpr$ times more (time-frequency) resources in comparison to single-packet reception case (i.e., the required number of resources is directly proportional to $\mpr$).
In effect, slots in $\mpr$-collision channel are $\mpr$ times larger compared to the standard (1-)collision channel, which is taken into account when assessing the performance, see Section~\ref{sec:model}.
This model is adequate for CDMA~\cite{MGA2017} or some $\mpr$-out-of-$\nuser$ coding schemes~\cite{GSP2018,MACbook,OP2017}.

Finally, we remark that the assumed model is in a certain sense conservative.
For instance, in non-orthogonal multiple-access schemes which rely on power-imbalances among users' transmissions, capture and successive interference cancellation, e.g.,~\cite{IXIT2014,CPMS2017}, the increase in time-frequency resources may not be needed.
In this respect, the results presented in this paper can be considered as a lower-bound on performance for the cases when $\mpr > 1$.

\section{System Model}
\label{sec:model}

Consider $\nuser$ active users and a common \ac{AP}.
The users are contending to access the \ac{AP} over a multiple-access $\mpr$-collision channel with feedback by transmitting fixed-length packets.
The time-frequency resources of the channel are divided into slots dimensioned to accommodate a single packet transmission. 
The users are synchronized on a slot basis via means of the feedback sent by the \ac{AP}.
The feedback channel is broadcast and assumed perfect. 


The contention starts with all $\nuser$ users transmitting in the first slot that appears on the channel and lasts until all users' packets are successfully decoded.
Henceforth, we also denote the event of successful decoding of a user packet as the user resolution.
In each slot, the contention outcome can be an idle (no user transmitted in the slot), collision (more than $\mpr$ users transmitted in the slot) or success (up to and including $\mpr$ users transmitted in the slot, and all their transmissions were successfully decoded).
In the last case, the success may trigger subsequent decodings via SIC along the tree, as depicted in Fig.~\ref{fig:tree-comparisons}c)\footnote{Recall that in the figure it is assumed that $\mpr=1$.}. 
All successfully decoded users do not contend anymore, while the collided users split into two groups.
We consider binary splitting, i.e., a user performing a split can join either a generic group 0 or group 1.
The choice is performed independently by each user, where group 0 is joined with probability $p$ and group 1 with probability $1-p$.
The users in group 0 contend in the next slot.
The users in group 1 wait - they can either become resolved through the application of SIC after a successful slot, as depicted in Fig.~\ref{fig:tree-comparisons}c) in the siblings of slots 4, 3 and 7, or they will perform an immediate split, which takes place after slot 6 and 4 in Fig.~\ref{fig:tree-comparisons}.
The decisions whether to transmit, split/or and wait are performed by monitoring the feedback channel and updating the state variables maintained by users.
For the sake of brevity, we will not elaborate these aspects here, but note that they can be derived using the well-know principles of tree-algorithms, e.g. see \cite{capetanakis1979tree,flajolet,SICTA}.

The period elapsed from the first slot up to and including the last slot in which all $\nuser$ users become resolved is denoted as \ac{CRI}.
The length of \ac{CRI} in slots conditioned on $\nuser$ is a random variable denoted by $\cri_\nuser$.
The basic performance parameter of interest is the expected value of $\cri_\nuser$, denoted by $\ecri_\nuser$, i.e., $ \ecri_\nuser = \mathrm{E}[ \cri_\nuser ]$.
Another important performance parameter is the conditional throughput
\begin{align}
\label{eq:T}
    \T_n = \frac{1}{\mpr}\frac{n}{\ecri_\nuser}.
\end{align}
The throughput in \eqref{eq:T} is the measure of the efficiency of resource use, where the normalization with $\mpr$ reflects the linear increase in resources required to achieve $\mpr$-\ac{MPR}.

The introduced \ac{CRP} is just a building block of a complete random access protocol.
Another block is a \ac{CAP} which regulates how the activated users access the channel, this way determining the number of users $n$ that enter the CRP.
The considered \ac{CRP} can be combined either with the gated \ac{CAP} or windowed \ac{CAP}, the details of which will be presented in Section~\ref{sec:arrivals}.

\section{Analysis}
\label{sec:analysis}

In this section, we analyze the performance of the introduced CRP, i.e., the expected conditional CRI length and the conditional throughput.

The conditional length of a \ac{CRI}, given $\nuser$ active users at its start, is\footnote{There is a benign misuse of notation in \eqref{eq:sicta_cases} for $\nuser > \mpr$, where both the random variables on the \ac{lhs} and \ac{rhs} are denoted by $l$, although the ones on the \ac{rhs} correspond to the CRI's that stem from splitting of the CRI that figures on the \ac{lhs}.}
\begin{align}
\label{eq:sicta_cases}
    \cri_\nuser = \begin{cases}
    1, & n = 0,1,\dots,\mpr \\
    \cri_{i} + \cri_{\nuser - i}, & n > \mpr
    \end{cases}
\end{align}
where $i$ and $n-i$ denote the number of users that chose group 0 and group 1, respectively.

The expected conditional length of \ac{CRI}, $\ecri_\nuser = \E \{ \cri_\nuser \}$, is thus
\begin{align}
\label{eq:SICTA_ecri}
    \ecri_\nuser = \begin{cases}
    1, & n = 0,1,\dots,\mpr \\
    \sum_{i = 0}^\nuser {\nuser \choose i} p^i ( 1 - p )^{\nuser - i } ( \ecri_i + \ecri_{\nuser - i }  ), &  \nuser > \mpr 
    \end{cases}
\end{align}
where $p$ is the probability of a user joining the first group.
By developing \eqref{eq:SICTA_ecri}, $\ecri_\nuser$ can be  calculated recursively through
\begin{align}
\label{eq:BTA_ecri_1}
    \ecri_\nuser = \begin{cases} 
    1, & \nuser \leq \mpr  \\
    \frac{p^n + (1- p)^n + 2 \sum_{i = 1}^{\nuser - 1} {n \choose i} p ^{\nuser - i} ( 1 - p )^{i } \ecri_i }{1- p^\nuser - ( 1 - p )^\nuser }, & \nuser > \mpr.
    \end{cases}
\end{align}

\subsection{Direct Expression for $\ecri_\nuser$}

For the derivation of the direct, i.e. non-recursive expression for $\ecri_\nuser$, we rely on the method that relies on generating functions, elaborated in~\cite{MF1985}.
We start by introducing the \ac{CPGF} of $\cri_\nuser$
\begin{align}
\pgf_\nuser(z) = \mathrm{E} \left\{ z^{\cri_n} \right\}
\end{align}
where, due to \eqref{eq:sicta_cases}, the following holds
\begin{align}
\label{eq:BTA_ic}
\pgf_0 (z) = \pgf_1(z) = \dots = \pgf_\mpr(z) = z.    
\end{align}
For $\nuser > \mpr$, we have
\begin{align}
\label{eq:BTA_cpgf}
\pgf_\nuser(z)  = \sum_{i=0}^\nuser  { \nuser \choose i } p^{i} ( 1 - p )^{\nuser - i} \pgf_{i}(z) \pgf_{\nuser - i}(z).
\end{align}
Note that the following holds
\begin{align}
\label{eq:new_1}
\ecri_\nuser = \frac{ d\pgf_\nuser ( z )}{d z } |_{z=1}.
\end{align}

The (unconditional) \ac{PGF} of \ac{CRI}, assuming that 
$\nuser$ obeys a Poisson distribution\footnote{This is an auxiliary assumption that will not limit the general nature of the derived results.} with a mean $x$, is given by
\begin{align}
& \pgf (x,z)  =  \sum_{\nuser = 0}^\infty \pgf_\nuser (z) \frac{x^\nuser}{\nuser!} e^{-x} \label{jpgf} \\
& =  \sum_{\nuser = 0}^\infty \frac{x^\nuser}{\nuser!} e^{-x} \sum_{i=0}^{\nuser} { \nuser \choose i } p^i ( 1 - p )^{\nuser - i} \pgf_{i}(z) \pgf_{\nuser - i}(z) +  \nonumber \\ 
           &   +\left(z - z^2\right) \sum_{k=0}^{\mpr} \frac{x^k}{k!} e^{-x} \label{upgf}
\end{align}
where we exploited \eqref{eq:BTA_ic}, \eqref{eq:BTA_cpgf}, and the fact that for $ n \leq \mpr $
\begin{align}
    \sum^\nuser_{i=0} {n \choose i} p^{\nuser - i} (1 - p)^i \pgf_{\nuser - i}(z) \pgf_{i}(z) = z^2.
\end{align}
The first term on \ac{rhs} in \eqref{upgf} can be transformed into
\begin{align}
\sum^\infty_{\nuser=0} \pgf_{\nuser}(z) \frac{\left( p x \right)^{\nuser}}{\nuser!} e^{-p x} \sum^\infty_{i = 0} & \pgf_{i}(z) \frac{\left( ( 1 - p ) x \right)^{i}}{i!} e^{-(1 - p ) x}
\end{align}
so that \eqref{upgf} becomes
\begin{align}
\label{upgf2}
\pgf(x,z) = \pgf(p x,z)\pgf\left( (1-p) x,z \right)  + \left(z-z^{2}\right)\sum_{k=0}^{\mpr} \frac{x^k}{k!} e^{-x}.
\end{align}

We introduce the \ac{TGF} of $\ecri_\nuser$ as 
\begin{align}
\label{TGF}
\ecri ( x ) = \frac{\partial \pgf ( x, z ) }{ \partial z}\Big|_{z=1} = \sum_{\nuser=0}^{\infty} \ecri_\nuser \frac{x^\nuser}{\nuser !} e^{-x}.
\end{align}
where we exploited \eqref{eq:new_1} and \eqref{jpgf} to obtain the last expression in \eqref{TGF}.

Taking the partial derivative of \eqref{upgf2} with respect to $z$ at $z=1$ yields
\begin{align}
\ecri (x) = \ecri (p x) +\ecri ((1-p) x) - \sum_{k=0}^{\mpr} \frac{x^k}{k!}  e^{-x}
\label{eq:step_2_n}
\end{align}
where we used the fact that $\pgf(x, 1) = 1$, $\forall x$.

In the next step, we assume the following power series representation of $\ecri (x)$
\begin{align}
\label{ps_ecri}
\ecri (x) = \sum^\infty_{\nuser=0} a_\nuser x^\nuser
\end{align}
where, using \eqref{eq:new_1}, it can be shown that
\begin{align}
\ecri_\nuser  = \sum^\nuser_{k=0} \frac{n!}{(\nuser-k)!}a_k.
\label{non_recursive}
\end{align}
We now compute $a_k$, $k = 0,1,\dots,\nuser $.
From \eqref{eq:BTA_ecri_1}, it follows that
\begin{align}
    a_k = \begin{cases}
    1, & k = 0 \\
    0, & k = 1,\dots, \mpr .
    \end{cases}
    \label{alpha_weight_t}
\end{align}
Substituting \eqref{ps_ecri} into \eqref{eq:step_2_n} and using Maclaurin series expansion for $e^{-x}$ yields
\begin{align}
 \sum^\infty_{\nuser=0} a_\nuser  ( 1 - p^\nuser - (1 - p)^\nuser) x^\nuser = & - \sum_{k=0}^{\mpr} \frac{x^k}{k!}\sum^\infty_{\nuser=0} (-1)^\nuser \frac{x^\nuser}{\nuser!} \nonumber \\
 = \sum_{\nuser = 0}^\infty  \sum_{k=0}^{\min(\nuser,\mpr)} & \frac{( - 1)^{\nuser-k+1}}{ k! (\nuser - k )!} x^\nuser . \label{eq:ps}  
\end{align}
For $\nuser \leq K$, it can be shown that  
\begin{align}
\sum_{k=0}^{\nuser} \frac{( - 1)^{\nuser-k+1}}{ k! (\nuser - k )!} =
\begin{cases}
-1, & n = 0 \\
0, & 0 < n \leq K
\end{cases}
\end{align}
which, coupled with \eqref{alpha_weight_t}, transforms \eqref{eq:ps} into
\begin{align}
 \sum^\infty_{\nuser=K+1} a_\nuser ( 1 - p^\nuser - & (1 - p)^\nuser) x^\nuser \nonumber = \\ 
 & = \sum_{\nuser = K+1}^\infty \sum_{k=0}^K \frac{( - 1)^{\nuser-k+1}}{ k! (\nuser - k )!} x^\nuser .  \label{eq:exp_coeffs}
\end{align}

\begin{figure}[t]
  \centering
  \includegraphics[width=0.65\columnwidth]{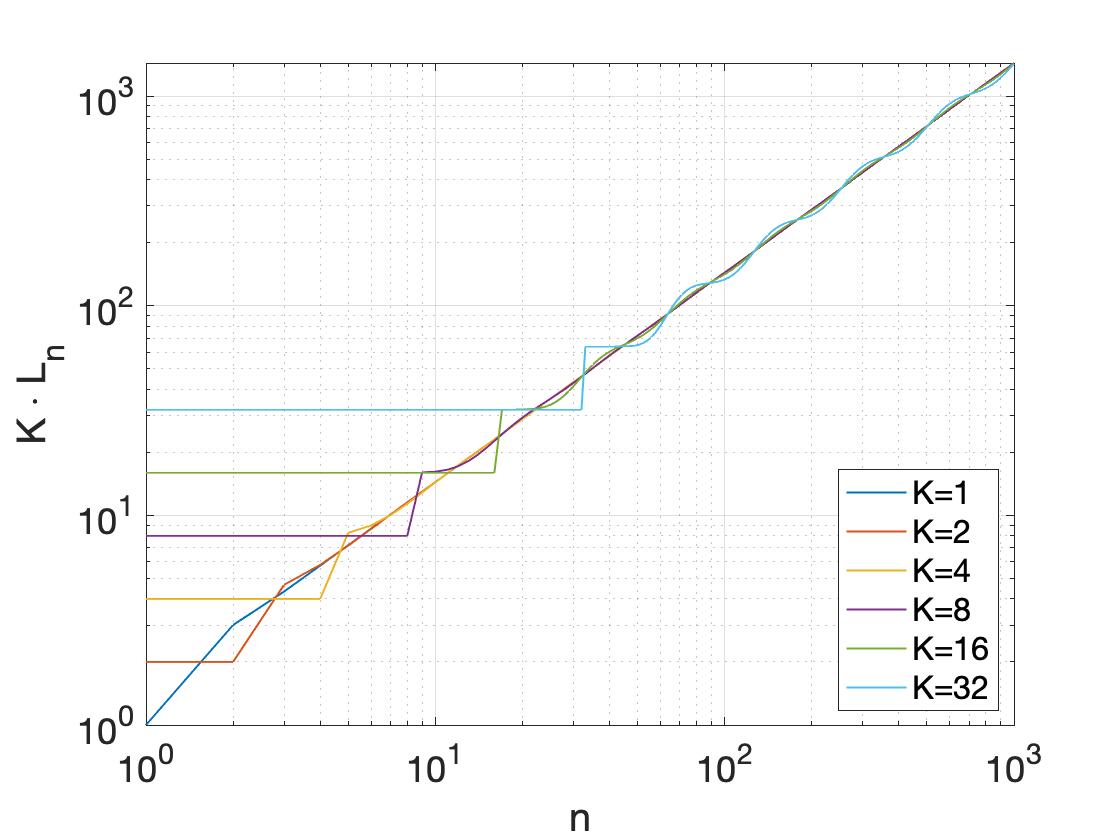}
  \caption{$K \cdot \ecri_\nuser$ as function of $n$ for $K \in \{ 1, 2, 4, 8, 16, 32 \}.$}
  \label{fig:L}
\end{figure}

Solving \eqref{eq:exp_coeffs} for $a_\nuser$, $\nuser \geq K$, we get
\begin{align}
a_\nuser  = {\sum_{k=0}^{\mpr} \frac{(-1)^{\nuser-k+1}}{k! (\nuser - k)!}}\cdot \frac{1}{1 - p^\nuser-(1-p)^\nuser}.
\label{alpha_weight_n}
\end{align}
By substituting \eqref{alpha_weight_t} and \eqref{alpha_weight_n} in \eqref{non_recursive} for $\nuser > \mpr$, and after some manipulation, we get
\begin{align}
\ecri_\nuser = 1 + \sum_{i = \mpr + 1}^\nuser {\nuser \choose i}  \frac{(-1)^{i+1}}{1 - p^i-(1-p)^i}\sum_{k=0}^{\mpr} {i \choose k} (-1)^{-k}.
\label{eq:mpr_bias_wosic}
\end{align}
Using the identity that holds for $ \mpr < j$
\begin{align}
    \sum_{k=0}^{\mpr} (-1)^{k} {i \choose k}  = ( - 1)^\mpr { i - 1 \choose \mpr }
\end{align}
\eqref{eq:mpr_bias_wosic} simplifies to 
\begin{align}
& \ecri_\nuser  = 1 + \sum_{i = \mpr + 1}^\nuser {\nuser \choose i}   {i-1 \choose \mpr} \frac{(-1)^{i -\mpr + 1} }{1 - p^i - ( 1 - p )^ i } \\
& = 1 +  {\nuser \choose \mpr} \sum_{i = \mpr + 1}^\nuser  { {n-\mpr \choose i-\mpr} \frac{ (i-\mpr)(-1)^{i-\mpr+1} }{i(1 - p^i-(1-p)^i)}}.
\end{align}
Finally, we get
\begin{align}
\label{eq:BTA_ecri_final}
\ecri_\nuser = 1 -  {\nuser \choose \mpr} \sum_{i = 1}^{\nuser - \mpr}  \frac{ i \, (-1)^{i} {\nuser -\mpr \choose i}} {(i + \mpr) ( 1 - p^{i+\mpr}-(1-p)^{i+\mpr})}.
\end{align}

\begin{figure}[t]
  \centering
  \includegraphics[width=0.65\columnwidth]{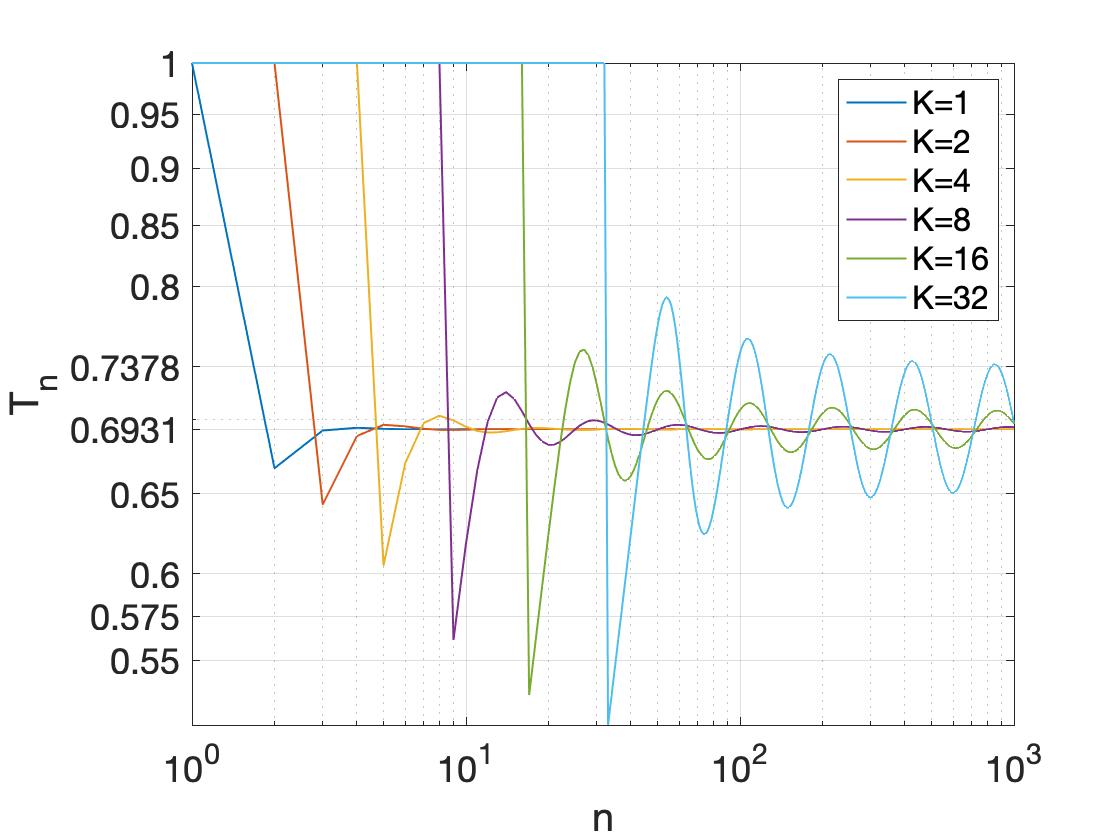}
  \caption{$\T_\nuser$ as function of $n$ for $K \in \{ 1, 2, 4, 8, 16, 32 \}.$}
  \label{fig:T}
\end{figure}

It is easy to show that \eqref{eq:BTA_ecri_final} is minimized for $p=\frac{1}{2}$.
In other words, fair splitting achieves minimal $\ecri_\nuser$, which is given by
\begin{align}
\label{eq:SICTA_ecri_fair}
\ecri_\nuser = 1 -  {\nuser \choose \mpr} \sum_{i = 1}^{\nuser - \mpr}  \frac{ i \, (-1)^{i} {\nuser -\mpr \choose i}} {(i + \mpr) ( 1 - 2^{-i-\mpr +1})}.
\end{align}       
In the rest of the paper, we assume fair splitting.

Fig.~\ref{fig:L} shows $\mpr \cdot \ecri_\nuser$, i.e., the expected conditional length of \ac{CRI} weighted by  $K$, to make the comparison fair among the curves obtained for different $\mpr$ (recall that it is assumed that the slot size increases linearly with $\mpr$).
Obviously, the curves for different $\mpr$ tend to each other as $\nuser$ increases. 
Also, a careful inspection can reveal that the curves show an oscillatory behaviour.
This oscillatory behaviour is more evident in Fig.~\ref{fig:T}, which shows the conditional throughput $\T_\nuser$ as function of $\nuser$.
The oscillations periodicity depending on $\log ( \nuser)$ and the oscillations amplitude increasing with $\mpr$. 
Further, the oscillations are non-vanishing, a fact identified in~\cite{MF1985} for the binary tree-algorithms on the standard collision channel.
We analytically investigate this phenomenon in the next subsection.

More importantly, both Fig.~\ref{fig:L} and Fig.~\ref{fig:T} suggest that the use of \ac{MPR} does not improve the performance of the tree-algorithm conditioned on $\nuser$, when normalized with $\mpr$.
In particular, Fig.~\ref{fig:T} shows that, as $\nuser \rightarrow \infty$, $\T_n$ oscillates around the value of $\ln (2) \approx 0.6931$, irrespective of the value of $\mpr$.
In Section~\ref{sec:arrivals}, we make further investigations of this issue. 

\subsection{Asymptotic behaviour of $\ecri_n$}
\label{sec:asymptotic}

Here we turn to analysis of asymptotic behavior of $\ecri_n$, enhancing the approach presented in \cite{MF1985}.
Rewriting \eqref{eq:step_2_n} for the case of fair-splitting (i.e., $p = 1/2$), we get
\begin{align}
\label{eq:f1}
    \ecri ( x ) - 2 \ecri \left( \frac{x}{2} \right) = - \sum_{k=0}^{\mpr} \frac{x^k}{k!}  e^{-x}.
\end{align}
By differentiating \eqref{eq:f1} twice, we get
\begin{align}
    \ecri''(x) - \frac{1}{2} \ecri''\left( \frac{x}{2} \right) = \left[\frac{x^{\mpr - 1 }}{( \mpr - 1 )!}  - \frac{x^{\mpr}}{\mpr!}\right] e^{-x} = g ( x )
\end{align}
which is a functional equation that satisfies the contraction condition and has the solution in the form~\cite{MF1985}
\begin{align}
\label{eq:f2}
    \ecri''(x) = & \sum_{m = 0}^{\infty} \frac{1}{2^m} \, g \left( \frac{x}{2^m} \right) \\
    = & \sum_{m = 0}^{\infty} \frac{1}{2^m} \left[ \frac{ \left( \frac{x}{2^m} \right)^{\mpr - 1 }}{( \mpr - 1 )!}  - \frac{ \left( \frac{x}{2^m} \right)^{\mpr}}{\mpr!} \right] e^{-\frac{x}{2^m}} .
\end{align}
Integrating \eqref{eq:f2} twice, and taking into account the initial conditions $ \ecri ( 0 ) = 1$ and $ \ecri' ( 0 )  = 0$ that stem from \eqref{TGF}, we obtain the following expression for the \ac{TGF}
\begin{align}
\label{eq:f3}
    \ecri ( x ) = 1 + \sum_{m = 0}^{\infty} 2^m - \sum_{m=0}^{\infty} 2^m e^{-\frac{x}{2^m}} \sum_{k = 0}^{\mpr} \frac{\left(\frac{x}{2^m}\right)^k }{k!}
\end{align}
Exploiting \eqref{TGF} further, the previous equation can be transformed to
\begin{align}
\label{eq:f4}
    \sum_{\nuser=0}^{\infty} \ecri_\nuser \frac{x^\nuser}{\nuser!} = e^{x} + e^x \sum_{m = 0}^{\infty} 2^m \left[  1 -  e^{-\frac{x}{2^m}} \sum_{k = 0}^{\mpr}  \frac{\left(\frac{x}{2^m}\right)^k }{k!} \right].
\end{align}
Using the Maclaurin series expansion for $e^x$, and after some manipulation, we transform \eqref{eq:f4} into
\begin{align}
    & \sum_{\nuser=0}^{\infty}  \ecri_\nuser \frac{x^\nuser}{\nuser!} = \sum_{\nuser=0}^{\infty} \frac{x^\nuser}{\nuser!} \times \nonumber \\
    & \left\{ 1 + \sum_{m=0}^{\infty} 
    2^m \left[ 1 - \sum_{k=0}^{\min\{ \nuser, \mpr \} } {\nuser \choose k } \frac{( 1 - \frac{1}{2^m})^{\nuser-k}}{2^{m k}}  \right] 
    \right\}.
\end{align}
Equating coefficients for $x^\nuser$, $ \nuser > \mpr$, we get
\begin{align}
\label{eq:f5}
    \ecri_\nuser = 1 + \sum_{m=0}^{\infty} 2^{m} \left[ 1 - \sum_{k=0}^{\mpr} { n \choose k} \frac{ \left( 1 - \frac{1}{2^m} \right)^{\nuser-k}}{2^{m k}} \right].
\end{align}
In principle, from \eqref{eq:f5} one can derive the same expression for $\ecri_\nuser$ given by \eqref{eq:SICTA_ecri_fair}.
However, we do not pursue this further.
Instead, assuming that $K$ is fixed, we exploit the following approximations for $\nuser \gg \mpr$
\begin{align}
    & \left ( 1 - \frac{1}{2^m} \right)^{\nuser - k}  = e^{-\frac{n}{2^m} ( 1 - \frac{k}{n}) + \left( \frac{n}{2^m} \right)^2 O( n^{-1})} \approx e^{ - \frac{n}{2^m}} \\
    & {\nuser \choose k } = \frac{n^k}{k!} \left( 1 - \frac{k (k-1)}{2} \Theta ( n{^-1} ) \right) \approx \frac{n^k}{k!}
\end{align}
which, substituted into \eqref{eq:f5}, yield
\begin{align}
\label{eq:f6}
    \ecri_\nuser \approx 1 + \sum_{m = 0}^{\infty} 2^m \left[ 1 - \sum_{k = 0}^{\mpr} \left( \frac{\nuser}{2^m} \right)^k \frac{e^{ - \frac{n}{2^m}}}{k!}\right].
\end{align}
Now, the task at hand is to isolate $n$ in \eqref{eq:f6}, such that summation over $m$ can be performed.
For this purpose, we exploit the method for the asymptotic analysis of harmonic sums~\cite{MF1985,flajolet}.
We introduce the following function
\begin{align}
    g ( x ) =  1 - \sum_{k = 0}^{\mpr} \frac{x^k}{k!}  e^{ - x}.
\end{align}
The Mellin transform of $g(x)$ is
\begin{align}
    G (s) & = \int_0^{\infty} g ( x ) x^{ s - 1 } dx = - \Gamma ( s ) \left[ 1 + \sum_{k=1}^K \frac{\prod_{i=0}^{k-1} (s + i)}{ k !} \right] \nonumber \\
    & = - ( s + 1 ) \Gamma ( s ) \left[ 1 + \frac{s}{2!} + s \sum_{k=3}^{K} \frac{\prod_{i = 2}^{k-1} ( s + i )} {k!} \right] \label{eq:mellin}
\end{align}
where $s$ is a complex variable laying in the fundamental strip (i.e., strip of convergence) given by $- 2 < \mathfrak{Re} (s) < 0$ and $\Gamma ( s )$ is the meromorphic extension of the Gamma function. 
The inverse Mellin transform for $x = n / 2^m$ is given by
\begin{align}
    \label{eq:inv_mellin}
     & g \left( \frac{n}{2^m} \right) = \frac{1}{2 \pi j}\int_{\eta - j \infty }^{\eta + j \infty} G (s) \left( \frac{n}{2^m} \right)^{-s} ds = \\
      & - \frac{1}{2 \pi j}\int_{\eta - j \infty }^{\eta + j \infty} \Gamma ( s ) \left[ 1 + \sum_{k=1}^K \frac{\prod_{i=0}^{k-1} (s + i)}{ k !} \right] \frac{n^{-s}}{2^{-ms}} ds \nonumber
\end{align}
where $\eta$ belongs to the fundamental strip.
Substituting \eqref{eq:inv_mellin} into \eqref{eq:f6}, and interchanging the order of summation and integration, we obtain
\begin{align}
\label{eq:D}
    \ecri_\nuser & \approx 1 + \frac{1}{2 \pi j} \int_{\eta - j \infty }^{\eta + j \infty} G (s) n^{-s}\sum_{m = 0}^{\infty} 2^{(s+1)m} \\
    & = 1 + \frac{1}{2 \pi j} \int_{\eta - j \infty }^{\eta + j \infty} \frac{G (s) n^{-s}}{1 - 2^{s+1}} ds.    \label{eq:f7}
\end{align}
The domain of absolute convergence of the series in \eqref{eq:D} is $\mathfrak{Re}(s) < -1 $.
Thus, the fundamental strip of the integrand
\begin{align}
   H (s ) = \frac{G (s) n^{-s} }{ 1 - 2^{s+1} }
\end{align}
lies in the intersection of the domain of absolute convergence of the series and the fundamental strip of $G(s)$, and is given by $-2 < \Re (s) < - 1$.
In this strip lies $\eta$ in \eqref{eq:f7}.

\begin{figure}[t]
  \centering
  \includegraphics[width=0.5\columnwidth]{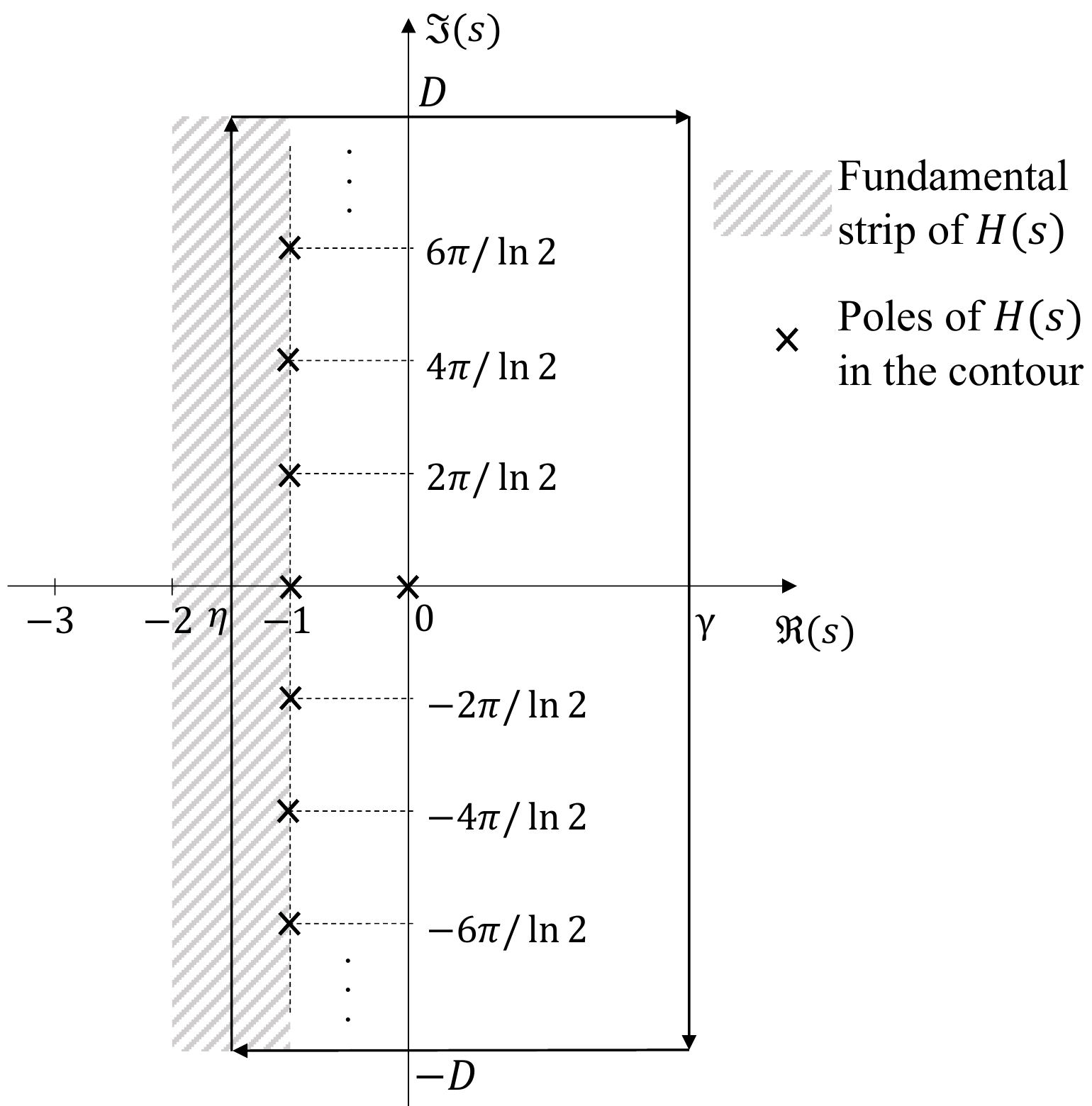}
  \caption{The contour of the integration in the complex plane.}
  \label{fig:contour}
\end{figure}

We compute the integral in \eqref{eq:f7} using the residue theorem. 
In order to evaluate $\ecri_\nuser$ for $\nuser \rightarrow \infty$, 
we set the set up the contour in the complex plane, depicted in Fig.~\ref{fig:contour}, where $\gamma$ and $D$ are some fixed values.
We close the path of integration in the half of the complex plane that is right to the fundamental strip, allowing $D$ and $\gamma$ to tend to infinity.
The  gamma  function  decays  exponentially  fast  as  the absolute value of the imaginary component of the argument increases, thus, the integration on the horizontal parts of the contour tends to zero as $|D| \rightarrow \infty$.
The integral on the vertical line $\Re ( s ) = \gamma$, $\gamma > 0$, is bounded by $O( n^{-\gamma})$, also tending to zero for large $n$~\cite[Chapter 5.2.2]{knuth}, \cite{flajolet}.
Thus, the integral in \eqref{eq:f7} is equal to the negative sum of the residues of the poles of $H (s)$ within the contour (negative due to the contour orientation).

The factor $n^{-s}$ trivially has no poles in the contour.
Further, $g ( s )$ has a simple pole in $0$ which is due to the corresponding pole of $\Gamma (s)$.\footnote{For the sake of completeness, we note that the pole of $\Gamma (s)$ at $-1$ is cancelled out by the zero $(s+1)$ of the function $G ( s )$, see \eqref{eq:mellin}.}
We have
\begin{align}
    \underset { s = 0 }{ \text{Res} } \, H ( s ) = - \Gamma ( 0 ) = - 1.
\end{align}

The factor $1 / (1 - 2^{s+1} )$ has simple poles at $s_p \in \mathcal{P} = \{ -1 + 2 \pi j m / \ln 2,  m \in \mathbb{Z} \}$, and it can be shown that the value of the corresponding residues is $ - 1 / \ln 2$. 
We first compute the value of the residue at $s_p=-1$
\begin{align}
    \underset{ s=-1 }{\mathrm{Res}} H(s) =  - \frac{G (-1) n}{ \ln 2} = \frac{n}{K \ln 2  }
\end{align}
where we used the fact that
\begin{align}
    G ( - 1 ) = - \left[ 1 - \sum_{k=2}^{K} \frac{(k-2)!}{k!} \right] = - \frac{1}{K}.
\end{align}
Further, for $s_p \in \{ -1 + 2\pi j m, m \in \mathbb{N} \}$, we have
\begin{align}
    \underset{ s = s_p }{\mathrm{Res}} & \, H(s) =  \nonumber \\
    - & \frac{1}{\ln 2} \Gamma \left( - 1 + \frac{2 \pi j m} { \ln 2} \right) n \, e^{ 2 \pi j m \log_2 n} A (K, m)
\end{align}
where
\begin{align}
\label{eq:A}
   A (K, m )  = 1 + \sum_{k = 1}^K \frac{\prod_{i=0}^{k-1} ( i - 1 + \frac{2 \pi j m }{ \ln 2}) }{k!}.
\end{align}
Similarly, for $s_p \in \{ -1 - 2\pi j m, m \in \mathbb{N} \}$ we have
\begin{align}
     \underset{ s = s_p }{\mathrm{Res}} & \, H(s) =  \\
    - & \frac{1}{\ln 2} \Gamma \left( - 1 - \frac{2 \pi j m} { \ln 2} \right) n \, e^{- 2 \pi j m \log_2 n} A (K, - m).
\end{align}
Using the mirror-symmetry property that holds for the gamma function
\begin{align}
    \Gamma ( s^*) = \Gamma^* ( s )
\end{align}
where $*$ denotes the complex conjugate, 
and using the following identity (which can be trivially shown)    
\begin{align}
    A ( K, - m ) = A^* ( K, m )  
\end{align}
we get
\begin{align}
    & \sum_{s_p \in \mathcal{P} \setminus \{0\} \} }  \underset{s = s_p} {\text{Res}} \, H ( s ) = - \frac{2n}{\ln 2} \sum_{m=1}^{\infty} \Re \left( B ( K, m ) e^{ 2 \pi j m \log_2 n } \right) \nonumber \\
    & = - \frac{2n}{\ln 2} \sum_{m=1}^{\infty} \left| B (K, m) \right| \cos \left( 2 \pi m \log_2 n + \arg \left( B (K, m) \right) \right) \label{eq:res1}
\end{align}
where
\begin{align}
    B ( K, m ) = \Gamma \left( - 1 + \frac{2 \pi j m}{\ln 2} \right) A (K, m ).
\end{align}
Again, since the gamma function decays exponentially fast as the imaginary component of the argument increases, \eqref{eq:res1} can be approximated as
\begin{align}
   \sum_{s_p \in \mathcal{P} \setminus \{0\} \} } & \underset{s = s_p} { \text{Res} \, H ( s )} \approx \label{eq:res2} \\
  & - \frac{2n}{\ln 2} \left| B (K, 1) \right| \cos \left( 2 \pi \log_2 n + \arg \left( B (K, 1) \right) \right). \nonumber
\end{align}
Putting all the pieces together, we obtain for the expected conditional length of CRI, when $\nuser \rightarrow \infty$, to be
\begin{align}
\label{eq:L_approx_final}
    \ecri_\nuser \approx & \frac{n}{K\ln 2} \times \nonumber\\
    & \left\{ 1 - 2 K | B ( K, 1 )|  \cos (2 \pi \log_2 n + \arg (B (K, 1 )) ) \right\}.
\end{align}
The conditional throughput, when $\nuser \rightarrow \infty$, is
\begin{align}
\label{eq:T_approx_final}
    \T_n & = \frac{n}{\mpr \ecri_n} \nonumber \\
    & \approx \frac{\ln 2}{ 1 - 2 K | B ( K, 1 )| \cos ( 2 \pi \log_2 n + \arg (B (K, 1)))}.
\end{align}
This oscillatory component in $\log_2 n$ was identified in, e.g.,~\cite{MF1985,flajolet,jong}.
In the case treated here, the difference is that its amplitude depends on  $K$ and can not be neglected, as it affects the stability bound (further discussed in Section~\ref{sec:arrivals}).

\begin{figure}[t]
  \centering
  \includegraphics[width=0.65\columnwidth]{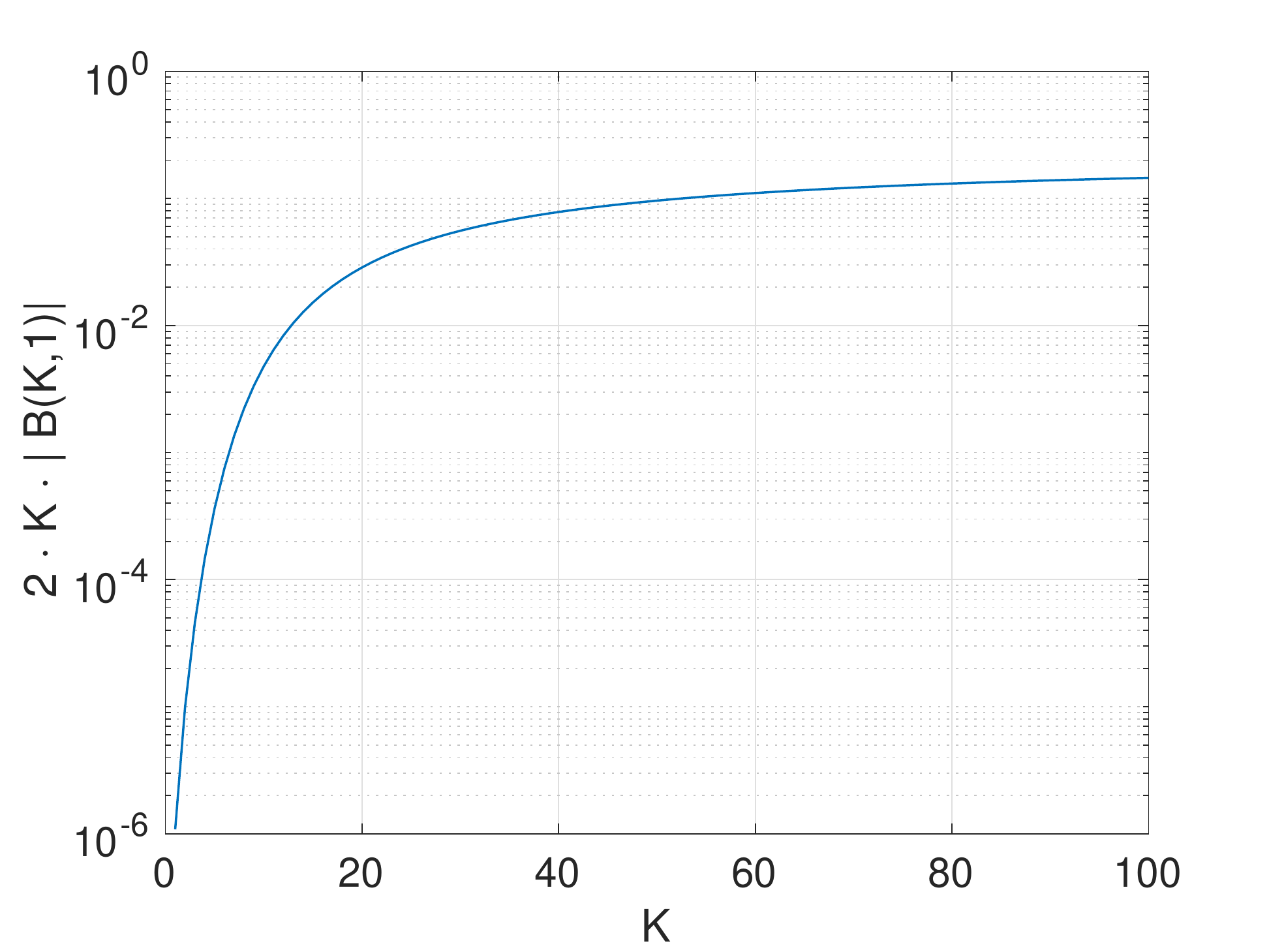}
  \caption{The amplitude of the oscillatory component in \eqref{eq:L_approx_final} and \eqref{eq:T_approx_final}, as function of $\mpr$.}
  \label{fig:2KBK}
\end{figure}

The expression $2K |B ( K, 1 )|$ can be easily computed for any $K$.
The graph presented in Fig.~\ref{fig:2KBK} shows that its value increases with $K$, which is also confirmed in Fig.~\ref{fig:L} and Fig.~\ref{fig:T}.
Although of a little practical relevance, an interesting problem in its own right is to determine the behaviour of $K |B ( K, 1 )|$ as $K \rightarrow \infty$. 
This problem is out of the paper scope; based on our preliminary investigation, we conjecture that that there is an upper bound on the value of $K |B ( K, 1 )|$ as $\mpr \rightarrow \infty$.

Finally, we validate the presented analysis by comparing its output with the results presented in Fig.~\ref{fig:T}.
For instance, $ 2 K |B ( K, 1 )|$ evaluates to 0.0607 
for $K=32$, implying that the asymptotic maximum and minimum values of $\T_\nuser$ are 0.7378 and 0.6536, respectively, see \eqref{eq:T_approx_final}.
Obviously, the curve for $T_n$ when $K=32$ in Fig.~\ref{fig:T} indeed tends to oscillate between these two values as $\nuser$ increases.






\subsection{Simple bounds on $\ecri_\nuser$ and $\T_\nuser$}
\label{sec:bounds}

We conclude this section by developing simple, but useful bounds on $\ecri_\nuser$ and $\T_\nuser$, that do not require asymptotic evaluation presented. 
In particular, these bounds can be computed for any finite $m$ and are valid for any $ n \geq m$. 

For $\nuser > K$ and fair splitting, the expected conditional length of \ac{CRI} reduces to
\begin{align}
\label{eq:fair_SICTA}
    \ecri_\nuser = \frac{\sum_{i = 0}^{\nuser - 1} {\nuser \choose i } \ecri_i }{ 2^{ \nuser - 1} - 1 }.
\end{align}

Building up on the method introduced by Massey~\cite{massey1981collision}, we want to find the constant $\alpha_m$ for which the following holds:
\begin{align}
\label{eq:cond}
    \ecri_\nuser \leq \alpha_m \nuser, \; \nuser \geq m.
\end{align}
For $n < m$, we can write
\begin{align}
\label{eq:n_leq_m}
    \ecri_\nuser \leq \alpha_m \nuser + \sum_{i = 1}^{M - 1} \delta_{i,\nuser}( \ecri_\nuser - \alpha_m \nuser )
\end{align}
where $\delta_{i,\nuser}$ is the Kronecker delta, and where \eqref{eq:n_leq_m} holds by definition.
In the induction step, we substitute \eqref{eq:n_leq_m} into \eqref{eq:fair_SICTA}, and after some manipulation, obtain
\begin{align}
\label{eq:bound_2}
    \ecri_\nuser \leq \alpha_m \nuser + \frac{ \sum_{i=0}^{m-1}{\nuser \choose i}(\ecri_i - \alpha_m i)}{2^{n - 1} -1 }
\end{align}
and the condition \eqref{eq:cond} will hold true for any
\begin{align}
    \alpha_m \geq \frac{\sum_{i=0}^{m-1} { \nuser \choose i} \ecri_i }{\sum_{i=0}^{m-1} { \nuser \choose i} i}
\end{align}
as the summation in the second term on the right-hand side of \eqref{eq:bound_2} is non-positive in this case.
The tightest upper bound is given by
\begin{align}
\label{eq:sup}
  \alpha_m = \sup_{\nuser \geq m} \frac{\sum_{i=0}^{m-1} { \nuser \choose i} \ecri_i }{\sum_{i=0}^{m-1} { \nuser \choose i} i}.   
\end{align}

\begin{table}[t]
  \begin{center}
    \caption{Bounds on expected conditional length of CRI and conditional throughput.}
    \label{tab:bounds}
    \begin{tabular}{|c||c|c|c|c|c|c|}
       \hline
       $\mpr$ &  $m$ & $n$ & $\alpha_m$ & $\beta_m$ & $ A_m $ & $B_m $  \\ \hline \hline
       1  & 50 & 100 & 1.4427 & 1.4427 &  0.6931  & 0.6931  \\ \hline
       2  & 100 & 200 & 0.7214 & 0.7213 & 0.6931 & 0.6932   \\ \hline
       4  & 200 & 400 & 0.3607 & 0.3606 & 0.6930 & 0.6933   \\ \hline
       8  & 400 & 800 & 0.1808 & 0.1799 & 0.6915 & 0.6948  \\ \hline
       16 & 400 & 800 & 0.0919 & 0.0884 & 0.6803 & 0.7069 \\ \hline
       32 & 400 & 800 & 0.0480 & 0.0421 & 0.6505 & 0.7420 \\ \hline
       64 & 500 & 1000 & 0.0254 & 0.0199 & 0.6141 & 0.7864 \\ \hline
    \end{tabular}
      \end{center}
\end{table}

In a completely analogous fashion, one can find the lower bound
 \begin{align}
     \ecri_\nuser \geq \beta_m n , \; \nuser \geq m 
 \end{align}
where
\begin{align}
\label{eq:inf}
\beta_m = \inf_{\nuser \geq m} \frac{\sum_{i=0}^{m-1} { \nuser \choose i} \ecri_i }{\sum_{i=0}^{m-1} { \nuser \choose i} i}. 
\end{align}
Note that the bounds in \eqref{eq:sup} and \eqref{eq:inf} can be made arbitrarily tight by increasing $m$ and $\nuser$.

The corresponding bounds on conditional throughput are simply
\begin{align}
    B_m = \frac{1}{\mpr \beta_m} \geq \T_\nuser \geq \frac{1}{\mpr \alpha_m} = A_m, \; \nuser \geq m.
\end{align}

In Table~\ref{tab:bounds}, we list $\alpha_m$, $\beta_m$, $A_m$ and $B_m$ (rounded up to four decimal places).
Note the agreement between the bounds on $\T_n$ shown in the table, i.e., $A_m$ and $B_m$, and the results plotted in Fig.~\ref{fig:T}.

\section{Performance Under Poisson Arrivals}
\label{sec:arrivals}

In this section, we provide insights into performance of a random access protocol that combines the \ac{CRP} protocol introduced in Section~\ref{sec:model} with the gated \ac{CAP} and the windowed \ac{CAP}.
We adopt the standard performance evaluation approach by assuming Poisson arrivals in an infinite user population;
the arrival intensity per slot is denoted by $\lambda$.
We are interested to identify the bounds on $\lambda$ for which the random access protocol features a stable operation.
In brief, the stability implies that the individual packets are successfully received with a finite delay almost surely~\cite{MF1985}. 

\subsection{Gated Access}
\label{sec:gated}

\begin{figure}[t]
  \centering
  \includegraphics[width=0.6\columnwidth]{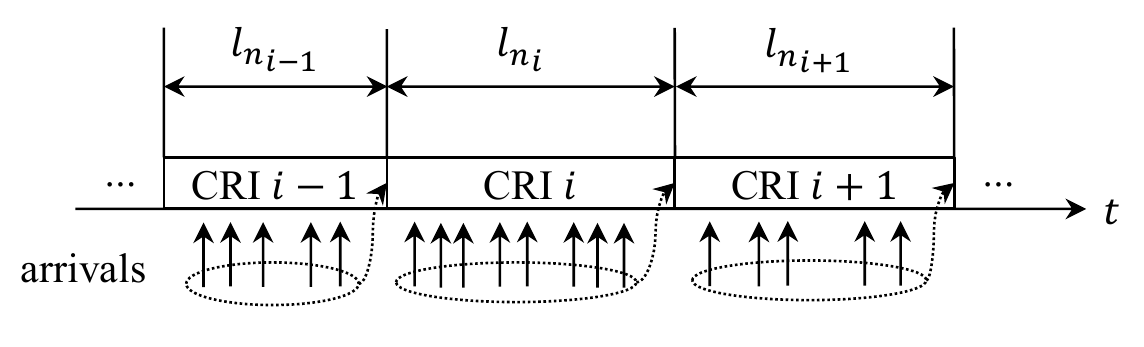}
  \caption{Illustration of the gated access: users arriving during $i$-th CRI are resolved in $(i+1)$-th CRI.}
  \label{fig:gated}
\end{figure}

The gated (also denoted as blocked) \ac{CAP} is an obvious approach to deal with traffic arrivals.
In particular, all users that arrive during a \ac{CRI} have to wait until that \ac{CRI} ends, i.e., they are blocked.
Once the current \ac{CRI} ends, all blocked users transmit in the next available slot, thus initiating the next \ac{CRI}.
Fig.~\ref{fig:gated} illustrates the principles of the gated access.

The stability conditions of the gated access were investigated in a number of works, e.g., in~\cite{MF1985,massey1981collision}.
The sufficient condition for stability is
\begin{align}
    \lambda < \lambda_\text{S} 
\end{align}
and the sufficient condition for instability is
\begin{align}
    \lambda > \lambda_\text{U} 
\end{align}
where the values of the bounds $\lambda_\text{S}$ and $\lambda_\text{U}$. 
Exploiting \eqref{eq:L_approx_final}, we get
\begin{align}
    \underset{\nuser \rightarrow \infty}{\lim \sup} \, \frac{\ecri_\nuser}{\nuser} = \frac{ 1 + 2 K | B( K, 1 )| }{K \ln (2)}  = \mathrm{L}_\text{S} \\
    \underset{\nuser \rightarrow \infty} {\lim \inf} \, \frac{\ecri_\nuser}{\nuser} = \frac{1 - 2 K | B( K, 1 )|}{K \ln (2)} = \mathrm{L}_\text{U}
\end{align}
from which it follows that~\cite{MF1985}
\begin{align}
    \lambda_\text{S} = \mathrm{L}_\text{S}^{-1} = \frac{K \ln (2)}{1 + 2 K | B( K, 1 )|} \\
    \lambda_\text{U} = \mathrm{L}_\text{U}^{-1} = \frac{K \ln (2)}{1 - 2 K | B( K, 1 )|}.
\end{align}

Table~\ref{tab:gated} lists values of $\lambda_\text{S} / \mpr$ and $\lambda_\text{U} / \mpr$ for several  values of $\mpr$; again, the normalization with $\mpr$ makes the comparison fair.\footnote{Note that $\lambda_\text{S} / \mpr = \underset{\nuser \rightarrow \infty}{\lim \inf} \, \T_n $ and $\lambda_\text{U} / \mpr = \underset{\nuser \rightarrow \infty}{\lim \sup} \, \T_n $, where $\T_n $ is given by \eqref{eq:T_approx_final}.}
Obviously, as $\mpr$ increases, difference among $\lambda_\text{S} / \mpr$ and $\lambda_\text{U} / \mpr$ grows.
This could be expected, since the amplitude of the oscillations in \eqref{eq:L_approx_final} grows with $\mpr$.


\begin{table}[t]
  \begin{center}
    \caption{Stability bounds on normalized traffic arrival intensity for gated access.}
    \label{tab:gated}
    \begin{tabular}{|c||c|c|}
       \hline
       $\mpr$ & $ \lambda_\text{S} / \mpr$ & $ \lambda_\text{U} / \mpr$ \\  \hline \hline
       1 & 0.6931 & 0.6931 \\ \hline
       2 & 0.6931 & 0.6932 \\ \hline
       4 & 0.6930 & 0.6932 \\ \hline
       8 &  0.6916 & 0.6947 \\ \hline
       16 & 0.6811 & 0.7056 \\ \hline
       32 & 0.6536 & 0.7378 \\ \hline
       64 & 0.6216 & 0.7833 \\ \hline
    \end{tabular}
      \end{center}
\end{table}


\subsection{Windowed Access}
\label{sec:windowed}

\begin{figure}[t]
  \centering
  \includegraphics[width=0.6\columnwidth]{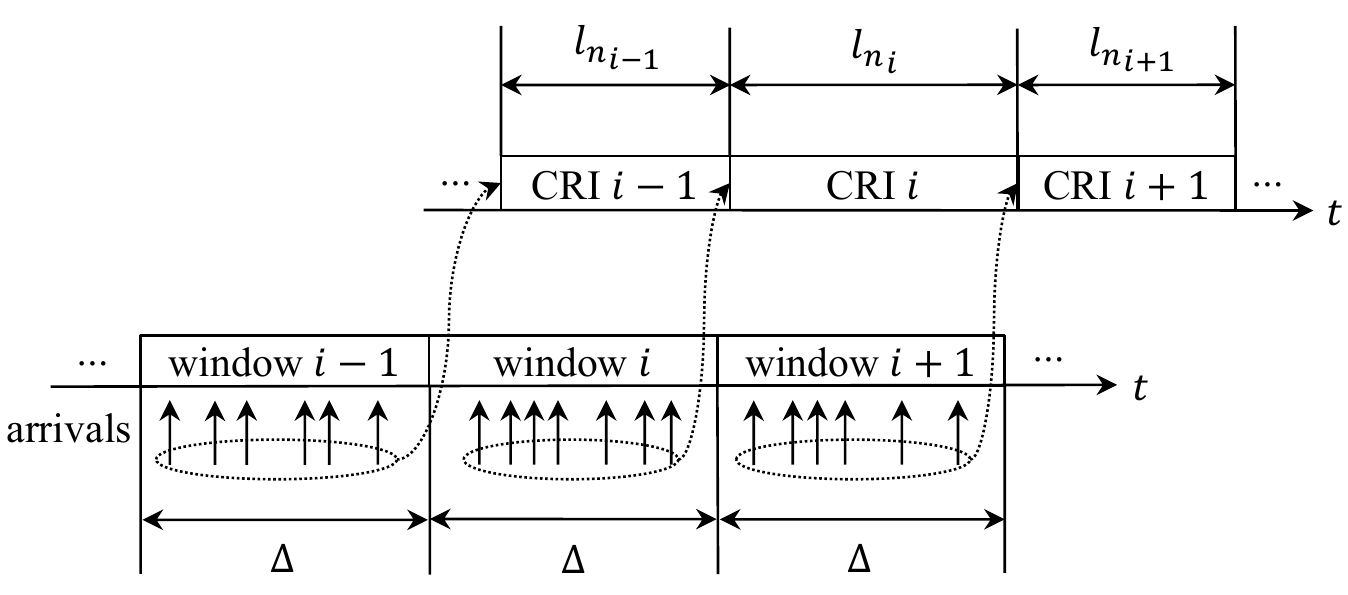}
  \caption{Illustration of the windowed access: users arriving during $i$-th window are grouped and resolved in a separate, corresponding CRI.}
  \label{fig:windowed}
\end{figure}

Another way to deal with the traffic arrivals is to use windowed \ac{CAP} (also denoted as the epoch mechanism).
In this approach, the time axis related to the traffic arrivals is divided into equal-length windows and every window is associated to a separate \ac{CRI}.
Specifically, the users arriving in $i$-th window transmit in the first slot after the \ac{CRI} of the users arriving in $(i-1)$-th window ends, thus starting their own \ac{CRI}.
Fig.~\ref{fig:windowed} illustrates the windowed access.


Denoting the window length in slots by $\Delta$ (which does not have to be an integer), the probability of $n$ arrivals ($\nuser \in \mathbb{N}$) in the window can be calculated as
\begin{align}
\label{eq:Poisson_window}
\Pr\{ N = \nuser \} = \frac{\lambda \Delta }{\nuser!}e^{ - \lambda \Delta},    
\end{align}
i.e., $n$ is a Poisson \ac{r.v.} with mean $\lambda \Delta$.
The expected length of \ac{CRI} is
\begin{align}
    \ecri(\lambda\Delta) = \E \{ \ecri_\nuser | \lambda \Delta \} = \sum_{\nuser=0}^\infty   \ecri_\nuser \frac{(\lambda\Delta)^\nuser}{\nuser!} e^{-\lambda\Delta}.
\label{eq:windowed_Acc}
\end{align}
The necessary condition for stability is the following
\begin{align}
\label{eq:WA_stability}
 \ecri(\lambda\Delta) < \Delta 
\end{align}
which is intuitively clear, as it ensures that arrivals in a window will be served (on average) in CRI that will last shorter than the window.\footnote{For stability to hold, the condition $\E \{ \ecri^2_\nuser \} < \infty$ has also to be satisfied. This can be shown for $\ecri(\lambda\Delta) < \Delta$, however, we omit the proof.} 

Exploiting the bounds derived in Section~\ref{sec:bounds}, it is easy to show that
\begin{align}
    f (\alpha_m, m, \lambda \Delta) \leq L ( \lambda \Delta ) \leq f (\beta_m, m, \lambda \Delta) 
\end{align}
where
\begin{align}
    f ( x, k, z ) = x \cdot z + \sum_{i = 0}^{k } ( L_i - x \cdot i ) \frac{ z^i }{i !} e^{ -z}.
\end{align}
The scheme will be stable if
\begin{align}
      f (\beta_m, m, \lambda\Delta) < \Delta 
\end{align}
which yields the bound on the arrival intensity
\begin{align}
    \label{eq:WA_SB}
    \lambda < \sup_{\lambda \Delta > 0 } \frac{\lambda \Delta}{f (\beta_m, m, \lambda\Delta)} = \lambda_\text{S}.
\end{align}
Similarly, the windowed access scheme will be unstable when
\begin{align}
\label{eq:WA_UB}
    \lambda > \sup_{\lambda \Delta > 0 } \frac{\lambda \Delta}{f (\alpha_m, m, \lambda\Delta)} = \lambda_\text{U}.
\end{align}

\begin{table}[t]
  \begin{center}
    \caption{Stability bounds on traffic arrival intensity for windowed access.}
    \label{tab:stable_throughput}
    \begin{tabular}{|c||c|c|c|}
       \hline
       $\mpr$ & $\lambda_\text{S} / \mpr$ & $\lambda_\text{U} / \mpr$ & $\lambda^*_\text{S} / \mpr$ (no SIC)~\cite{SGDK2020} \\ \hline \hline
       1 & 0.6931 & 0.6931 & 0.423 \\ \hline
       2 & 0.6932 & 0.6932 & 0.4707 \\ \hline
       4 & 0.6932 & 0.6932 & 0.5175 \\ \hline
       8 & 0.6947 & 0.6947 & 0.5678 \\ \hline
       16 & 0.7056 & 0.7056 & 0.6239 \\ \hline
       32 & 0.737 & 0.737 &  0.6862 \\ \hline
       64 & 0.7816 & 0.7816 &  0.7475 \\ \hline
    \end{tabular}
      \end{center}
\end{table}

Table~\ref{tab:stable_throughput} lists values of $\lambda_\text{S} / \mpr$ and $\lambda_\text{U} / \mpr$, calculated using the values of $\alpha_m$ and $\beta_m$ given in Table~\ref{tab:bounds}.
Evidently, a tangible increase in the maximum normalized arrival intensity $\lambda_\text{S} / \mpr$ for which the windowed scheme features a stable operation requires a substantial increase in $K$.
The table also shows the maximum normalized arrival intensity  $\lambda^*_\text{S} / \mpr$ for which the $K$-MPR BTA with windowed access~\cite{SGDK2020} and without SIC has a stable operation. 
The comparison between $\lambda_\text{S} / \mpr$ and $\lambda^*_\text{S} / \mpr$ reveals that, as $\mpr$ increases, most of the gain comes from the \ac{MPR}, while the contribution of \ac{SIC} becomes limited.

\begin{figure}[t]
  \centering
  \includegraphics[width=0.65\columnwidth]{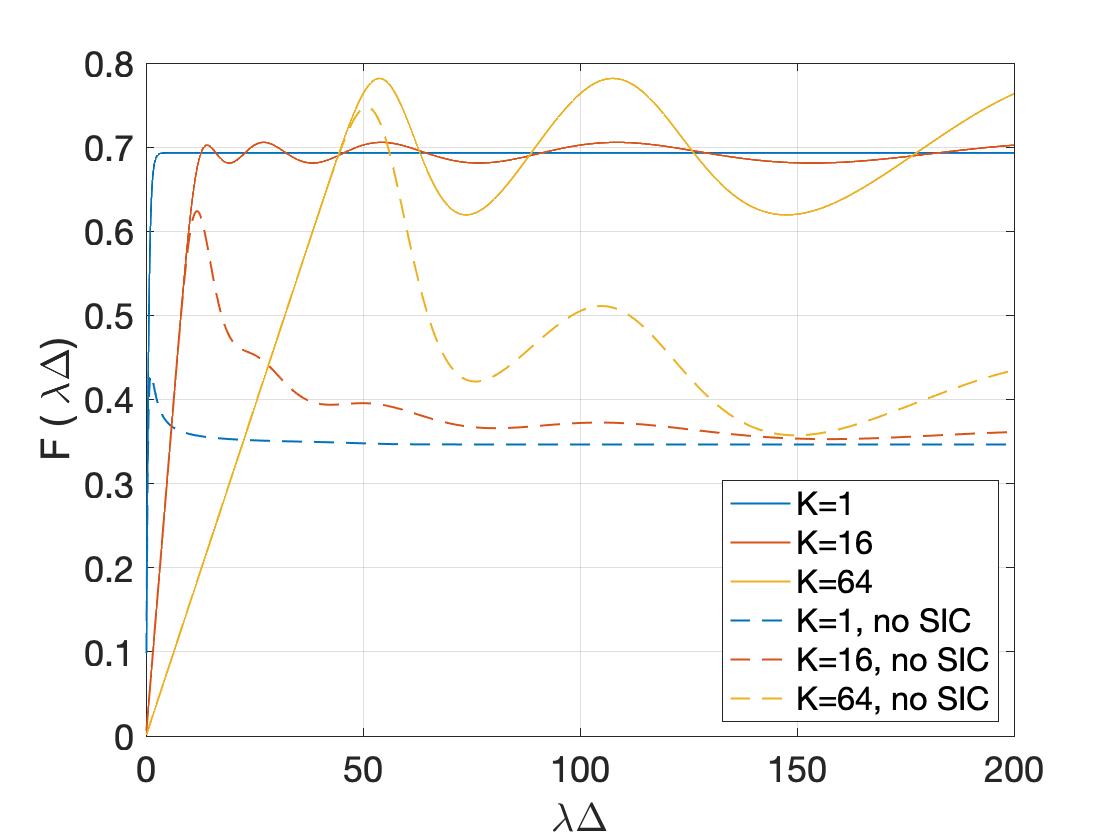}
  \caption{Sensitivity of the stability bound on the arrival intensity within a window as function of the average number of arrivals in the window $\lambda \Delta$.}
  \label{fig:sensitivity}
\end{figure}

In Fig.~\ref{fig:sensitivity}, we plot $F ( \lambda \Delta) = \frac{\lambda \Delta}{\mpr f (\beta_m, m, \lambda \Delta)}$, see \eqref{eq:WA_SB}, as function of $\lambda \Delta$, i.e., the sensitivity of the stability bound on the normalized arrival intensity per slot as function of the arrival intensity within a window.
The figure shows the characteristic oscillatory behaviour, which becomes more pronounced as $\mpr$ increases.
Nevertheless, the oscillations' periodicity is rather large, implying that there is a certain tolerance on the potential estimation errors of $\lambda$ and/or dimensioning errors of window length $\Delta$.
We also plot the sensitivity of the stability bound for the analogous protocol without SIC, investigated in~\cite{SGDK2020}.
Obviously, as $\mpr$ increases, the bound has a clearly pronounced maximum, and after which the performance quickly deteriorates. 
It can be concluded that in this respect, the protocol with SIC is an advantageous solution.


\section{Discussion and Conclusions}
\label{sec:conclusion}

The method for the derivation of the asymptotic values of the expected conditional CRI length and the throughput presented in Section~\ref{sec:asymptotic} can be extended to the case of \ac{BTA} with fair splitting and $\mpr$-\ac{MPR} (without SIC)~\cite{SGDK2020}.
Here we give the final expressions without a formal proof.
Specifically, for $\nuser \geq \mpr$, the expected conditional CRI length is
\begin{align}
    \ecri^*_\nuser = 1 + 2 \sum_{m=0}^{\infty} 2^{m} \left[ 1 - \sum_{k=0}^{\mpr} { n \choose k} \frac{ \left( 1 - \frac{1}{2^m} \right)^{\nuser-k}}{2^{m k}} \right]
\end{align}
which is asymptotically
\begin{align}
    \ecri^*_\nuser & \approx 1 + 2 \sum_{m = 0}^{\infty} 2^m \left[ 1 - \sum_{k = 0}^{\mpr} \left( \frac{\nuser}{2^m} \right)^k \frac{e^{ - \frac{n}{2^m}}}{k!}\right]. 
\end{align}   
Using the Mellin-transform based asymptotic analysis, we get
\begin{align}   
    \ecri^*_\nuser & \approx - 1 + \frac{2n}{K\ln 2} \times \nonumber \\
    & \left( 1 - 2 K | B ( K, 1 )| \right) \cos (2 \pi \log_2 n + \arg (B (K, 1 )) ) \\
    & \approx \frac{2n}{K\ln 2} \times \nonumber \\
    & \left( 1 - 2 K | B ( K, 1 )| \right) \cos (2 \pi \log_2 n + \arg (B (K, 1 )) ).
\end{align}
The conditional throughout is then simply
\begin{align}
    \T^*_\nuser \approx \frac{\ln 2}{ 2 \left[ 1 - 2 K | B ( K, 1 )| \cos \left( 2 \pi \log_2 n + \arg \left( B (K, 1) \right) \right) \right] }.
\end{align}
The last expression confirms the result identified in~\cite{SGDK2020}, that the conditional throughout of BTA with $\mpr$-\ac{MPR} oscillates around the value of $\ln 2 / 2 \approx 0.347$ as $\nuser \rightarrow \infty$.

We now turn to the comparison between the considered scheme with analogous schemes from the slotted ALOHA family that exploit \ac{MPR} and \ac{SIC}.
We mention \ac{IRSA}~\cite{L2011}, a frame slotted ALOHA protocol in which active users transmit several replicas of their packets in the frame.
Asymptotically, \ac{IRSA} supports load thresholds $G^*$ (defined as the ratio of the number of users and slots in the frame) close to 1 with the user resolution probability tending to 1, when the number of replicas transmitted per user is drawn according to a predefined, optimized distribution.
Essentially, this performance parameter is equivalent to the throughput.
In~\cite{GS2013} it was shown that, when \ac{IRSA} is coupled with \ac{MPR}, the normalized load threshold $G^* / \mpr$ for a fixed maximum number of replicas per user, decreases with $\mpr$ when $\nuser \rightarrow \infty$.
A similar insight, in terms of the upper bound on $G^* / \mpr$ was shown in~\cite{HAMK2020}.
On the other hand, the work in~\cite{SPL2017} showed that for the generalized variant of \ac{IRSA}, denoted as \ac{CSA}, the converse bound on $G^* / \mpr$ increases with $K$, quickly becoming very close to 1, and that, asymptotically, spatially-coupled \ac{CSA} operates close to the bound, which is out of the reach of the scheme considered in this paper.
This evidence may lead to a conclusion that \ac{IRSA} based schemes represent a better choice.
However, for a finite number of contending users, the maximum normalized load which the probability of successful user resolution is close to 1 in \ac{IRSA} is $G / \mpr \lesssim 0.7$~\cite{HAMK2020},\footnote{Also growing with $\mpr$, in contrast to the asymptotic behaviour.} which is comparable to the \ac{MST} of the scheme considered in this paper.
It should also be noted that slotted ALOHA-based protocols, in general, require some form of stabilization, while \ac{IRSA}-like protocols additionally require (i) optimization of the frame length, (ii) optimization of the distribution that governs the choice of the number of replicas, and (iii) placement of the pointers in the packet headers, required for the \ac{SIC}.
In contrast, tree-protocols are inherently stable for loads up to the \ac{MST}, and in case of the windowed access, require just the optimization of the window length.
Thus, for systems that support a frequent feedback (i.e., after every uplink slot), tree-algorithms with \ac{MPR} and \ac{SIC} may represent a suitable random-access solution.

Finally, we comment on an approach through which the performance of the scheme could be pushed further.
Specifically, as shown in~\cite{SSP2013}, one of the factors limiting the performance of tree algorithms with \ac{SIC} is a too high fraction of singleton slots in comparison to \ac{IRSA}-like protocols, which are unavoidable due to the very nature of the collision resolution process.
A way to address this drawback is to form a set of partially-split trees pertaining to the same initial collision and perform \ac{SIC} over the whole set.
It remains to be seen how the addition of \ac{MPR} to the framework would affect the performance of such scheme.






\bibliographystyle{IEEEtran}

\bibliography{bibliography}

\begin{thebibliography}{10}
\providecommand{\url}[1]{#1}
\csname url@samestyle\endcsname
\providecommand{\newblock}{\relax}
\providecommand{\bibinfo}[2]{#2}
\providecommand{\BIBentrySTDinterwordspacing}{\spaceskip=0pt\relax}
\providecommand{\BIBentryALTinterwordstretchfactor}{4}
\providecommand{\BIBentryALTinterwordspacing}{\spaceskip=\fontdimen2\font plus
\BIBentryALTinterwordstretchfactor\fontdimen3\font minus
  \fontdimen4\font\relax}
\providecommand{\BIBforeignlanguage}[2]{{%
\expandafter\ifx\csname l@#1\endcsname\relax
\typeout{** WARNING: IEEEtran.bst: No hyphenation pattern has been}%
\typeout{** loaded for the language `#1'. Using the pattern for}%
\typeout{** the default language instead.}%
\else
\language=\csname l@#1\endcsname
\fi
#2}}
\providecommand{\BIBdecl}{\relax}
\BIBdecl

\bibitem{capetanakis1979tree}
J.~Capetanakis, ``Tree algorithms for packet broadcast channels,'' \emph{IEEE
  Trans. Info. Theory}, vol.~25, no.~5, pp. 505--515, Sep. 1979.

\bibitem{Abramson:ALOHA}
N.~Abramson, ``The {ALOHA} system -- {A}nother alternative for computer
  communications,'' in \emph{Proc. of 1970 Fall Joint Computer Conf.},
  vol.~37.\hskip 1em plus 0.5em minus 0.4em\relax AFIPS Press, 1970, pp.
  281--285.

\bibitem{R1975}
L.~G. Roberts, ``{ALOHA} packet system with and without slots and capture,''
  \emph{SIGCOMM Comput. Commun. Rev.}, vol.~5, no.~2, pp. 28--42, Apr. 1975.

\bibitem{GVS1988}
S.~Ghez, S.~Verdu, and S.~C. Schwartz, ``{S}tability {P}roperties of {S}lotted
  {ALOHA} with {M}ultipacket {R}eception {C}apability,'' \emph{IEEE Trans.
  Autom. Control}, vol.~33, no.~7, pp. 640--649, Jul. 1988.

\bibitem{ZZ2012}
A.~Zanella and M.~Zorzi, ``{T}heoretical {A}nalysis of the {C}apture
  {P}robability in {W}ireless {S}ystems with {M}ultiple {P}acket {R}eception
  {C}apabilities,'' \emph{IEEE Trans. Commun.}, vol.~60, no.~4, pp. 1058--1071,
  Apr. 2012.

\bibitem{GSP2015}
J.~Goseling, C.~Stefanovic, and P.~Popovski, ``{A} {P}seudo-{B}ayesian
  {A}pproach to {S}ign-{C}ompute-{R}esolve {S}lotted {ALOHA},'' in \emph{Proc.
  IEEE ICC 2015, MASSAP Workshop}, London, UK, Jun. 2015.

\bibitem{PSLP2014}
E.~Paolini, C.~Stefanovic, G.~Liva, and P.~Popovski, ``Coded random access:
  {H}ow coding theory helps to build random access protocols,'' \emph{IEEE
  Commun. Mag.}, vol.~53, no.~6, pp. 144--150, Jun. 2015.

\bibitem{L2011}
G.~Liva, ``Graph-based analysis and optimization of contention resolution
  diversity slotted {ALOHA},'' \emph{IEEE Trans. Commun.}, vol.~59, no.~2, pp.
  477--487, Feb. 2011.

\bibitem{Liva2015:CodedAloha}
E.~Paolini, G.~Liva, and M.~Chiani, ``{C}oded slotted {ALOHA}: {A} graph-based
  method for uncoordinated multiple access,'' \emph{IEEE Trans. Info. Theory},
  vol.~61, no.~12, pp. 6815--6832, Dec. 2015.

\bibitem{SPL2017}
C.~Stefanovic, E.~Paolini, and G.~Liva, ``{Asymptotic Performance of Coded
  Slotted ALOHA with Multi Packet Reception},'' \emph{IEEE Commun. Lett.},
  vol.~22, no.~1, pp. 105--108, Jan. 2018.

\bibitem{SICTA}
Y.~Yu and G.~B. Giannakis, ``{High-Throughput Random Access Using Successive
  Interference Cancellation in a Tree Algorithm},'' \emph{IEEE Trans. Info.
  Theory}, vol.~53, no.~12, pp. 4628--4639, Dec. 2007.

\bibitem{V1986}
S.~Verdu, ``Computation of the efficiency of the mosely-humblet contention
  resolution algorithm: A simple method,'' \emph{Proceedings of the IEEE},
  vol.~74, no.~4, pp. 613--614, Apr. 1986.

\bibitem{SGDK2020}
C.~Stefanovic, H.~M. Gürsu, Y.~Deshpande, and W.~Kellerer, ``{Analysis of
  Tree-Algorithms with Multi-Packet Reception},'' in \emph{Proc. IEEE GLOBECOM
  2020}, Taipei, Taiwan, Dec. 2020.

\bibitem{massey1981collision}
J.~L. Massey, ``Collision-resolution algorithms and random-access
  communications,'' in \emph{Multi-user communication systems}.\hskip 1em plus
  0.5em minus 0.4em\relax Springer, 1981, pp. 73--137.

\bibitem{MF1985}
P.~Mathys and P.~Flajolet, ``{Q}-ary {C}ollision {R}esolution {A}lgorithms in
  {R}andom-{A}ccess {S}ystems with {F}ree or {B}locked {C}hannel {A}ccess,''
  \emph{IEEE Trans. Info. Theory}, vol.~31, no.~2, pp. 217--243, Mar. 1985.

\bibitem{TsyMikLik83}
B.~S. Tsybakov, V.~A. Mikhailov, and N.~B. Likhanov, ``{Bounds for Packet
  Transmission Rate in a~Random-Multiple-Access System},'' \emph{Probl.
  Peredachi Inf.}, vol.~19, no.~1, pp. 61--81, 1983.

\bibitem{LikPloSha93}
N.~B. Likhanov, I.~Plotnik, Y.~Shavitt, M.~Sidi, and B.~S. Tsybakov, ``{Random
  Access Algorithms with Multiple Reception Capability and $n$-ary Feedback
  Channel},'' \emph{Probl. Peredachi Inf.}, vol.~29, no.~1, pp. 82--91, 1993.

\bibitem{gau2011tree}
R.-H. Gau, ``Tree/stack splitting with remainder for distributed wireless
  medium access control with multipacket reception,'' \emph{IEEE Trans. Wirel.
  Commun.}, vol.~10, no.~11, pp. 3909--3923, Nov. 2011.

\bibitem{d-ary}
\BIBentryALTinterwordspacing
Y.~Deshpande, C.~Stefanovic, H.~M. Gürsu, and W.~Kellerer, ``On d-ary tree
  algorithms with successive interference cancellation,'' 2022. [Online].
  Available: \url{https://arxiv.org/abs/2208.04066}
\BIBentrySTDinterwordspacing

\bibitem{GSP2018}
J.~{Goseling}, C.~{Stefanovic}, and P.~{Popovski}, ``{Sign-Compute-Resolve for
  Tree Splitting Random Access},'' \emph{{IEEE} Trans. Inf. Theory}, vol.~64,
  no.~7, pp. 5261--5276, Jul. 2018.

\bibitem{PB2009}
G.~T. Peeters and B.~Van~Houdt, ``{On the Maximum Stable Throughput of Tree
  Algorithms With Free Access},'' \emph{IEEE Trans. Info. Theory}, vol.~55,
  no.~11, pp. 5087--5099, Nov. 2009.

\bibitem{3GPPTS36.321}
3GPP, ``{TS36.321 v16.3.0} - {M}edium {A}ccess {C}ontrol {(MAC)} protocol
  specification ({R}elease 16).'' Tech. Rep., Dec. 2020.

\bibitem{macbook-original}
E.~Biglieri and L.~Gyorfi, Eds., \emph{{Multiple Access Channels}}.\hskip 1em
  plus 0.5em minus 0.4em\relax IOS press, 2007.

\bibitem{OP2017}
O.~Ordentlich and Y.~Polyanskiy, ``Low complexity schemes for the random access
  gaussian channel,'' in \emph{Proc. IEEE ISIT 2017}, Jun. 2017, pp.
  2528--2532.

\bibitem{Ghez1988}
S.~Ghez, S.~Verd{\'u}, and S.~Schwartz, ``{Stability Properties of Slotted
  {ALOHA} with Multipacket Reception Capability},'' \emph{IEEE Trans. Automat.
  Contr.}, vol.~33, no.~7, pp. 640--649, Jul. 1988.

\bibitem{TZM2001}
L.~Tong, Q.~Zhao, and G.~Mergen, ``Multipacket reception in random access
  wireless networks: from signal processing to optimal medium access control,''
  \emph{IEEE Commun. Maga.}, vol.~39, no.~11, pp. 108--112, Nov. 2001.

\bibitem{MGA2017}
A.~{Mengali}, R.~{De Gaudenzi}, and P.~{Arapoglou}, ``Enhancing the physical
  layer of contention resolution diversity slotted {ALOHA},'' \emph{{IEEE
  Trans. Commun.}}, vol.~65, no.~10, pp. 4295--4308, Oct. 2017.

\bibitem{MACbook}
D.~Danyev, B.~Laczay, and M.~Ruszinko,
  ``\BIBforeignlanguage{English}{{M}ultiple {A}ccess {A}dder {C}hannel},'' in
  \emph{\BIBforeignlanguage{English}{Multiple Access Channels}}, E.~Biglieri
  and L.~Gyorfi, Eds.\hskip 1em plus 0.5em minus 0.4em\relax IOS press, 2007,
  pp. 26--53.

\bibitem{IXIT2014}
M.~Al-Imari, P.~Xiao, M.~A. Imran, and R.~Tafazolli, ``{Uplink non-orthogonal
  multiple access for 5G wireless networks},'' in \emph{Proc. IEEE ISWCS 2014},
  2014.

\bibitem{CPMS2017}
F.~Clazzer, E.~Paolini, I.~Mambelli, and C.~Stefanovic, ``{Irregular repetition
  slotted ALOHA over the Rayleigh block fading channel with capture},'' in
  \emph{Proc. IEEE ICC 2017}, May 2017, pp. 1--6.

\bibitem{flajolet}
P.~Flajolet, X.~Gourdon, and P.~Dumas, ``Mellin transforms and asymptotics:
  Harmonic sums,'' \emph{Theor. Comput. Sci.}, vol. 144, pp. 3--58, 1995.

\bibitem{knuth}
D.~E. Knuth, \emph{{The Art of Computer Programming, 2nd ed.}}\hskip 1em plus
  0.5em minus 0.4em\relax Addison - Wesley, 1998, vol.~3.

\bibitem{jong}
A.~Janssen and M.~de~Jong, ``Analysis of contention tree algorithms,''
  \emph{IEEE Trans. Info. Theory}, vol.~46, no.~6, pp. 2163--2172, 2000.

\bibitem{GS2013}
M.~Ghanbarinejad and C.~Schlegel, ``{Irregular Repetition Slotted ALOHA with
  Multiuser Detection},'' in \emph{Proc. IEEE WONS 2013}, Banff, AB, Canada,
  Mar. 2013.

\bibitem{HAMK2020}
I.~Hmedoush, C.~Adjih, P.~Mühlethaler, and V.~Kumar, ``{On the Performance of
  Irregular Repetition Slotted Aloha with Multiple Packet Reception},'' in
  \emph{Proc. IEEE IWCMC 2020}, 2020, pp. 557--564.

\bibitem{SSP2013}
J.~H. Sørensen, C.~Stefanovic, and P.~Popovski, ``Coded splitting tree
  protocols,'' in \emph{Proc. IEEE ISIT 2013}, 2013, pp. 2860--2864.

\end{thebibliography}

\end{document}